\documentclass[]{aastex631}

\usepackage{amsmath}

\newcommand{\Ha}{H\ensuremath{\alpha}}

\definecolor{Red}{rgb}{1,0,0}

\definecolor{terry}{rgb}{0.0,0.5,0.0}

\submitjournal{ApJ}
\accepted{January 6, 2023}
\shorttitle{Nonlinear Fast Waves in a Prominence Pillar}
\shortauthors{Ofman et al.}

\begin{document}

\title{Nonlinear Fast Magnetosonic Waves in Solar Prominence Pillars}

\correspondingauthor{Leon Ofman}
\email{ofman@cua.edu}
\author[0000-0003-0602-6693]{Leon Ofman}
\affiliation{Department of Physics\\
Catholic University of America  \\
Washington, DC 20064, USA}
\affiliation{Heliophysics Science Division \\
NASA Goddard Space Flight Center \\
Greenbelt, MD 20771, USA}
\altaffiliation{Visiting, Department of Geosciences,  Tel Aviv University, Tel Aviv, Israel}

\author{Therese A.\ Kucera}
\affiliation{Heliophysics Science Division \\
NASA Goddard Space Flight Center \\
Greenbelt, MD 20771, USA}

\author{C.\ Richard DeVore}
\affiliation{Heliophysics Science Division \\
NASA Goddard Space Flight Center \\
Greenbelt, MD 20771, USA}




\begin{abstract}

We investigate the properties of nonlinear fast magnetosonic (NFM) waves in a solar prominence, motivated by recent high-resolution and high-cadence Hinode/SOT observations of small-scale oscillations in a prominence pillar. As an example, we  analyze the details of the 2012 February 14 Hinode/SOT observations of quasi-periodic propagating features consistent with NFM waves, imaged in emission in Ca~II and in the far blue wing of H$\alpha$. We perform wavelet analysis and find oscillations in the 1-3 min period range. Guided by these observations, we model the NFM waves with a three-dimensional magnetohydrodynamics (3D MHD) model, extending previous 2.5D MHD studies. The new model includes the structure of the high-density, low-temperature material of the prominence pillar embedded in the hot corona, in both potential and non-force-free sheared magnetic field configurations. The nonlinear model demonstrates the effects of mode coupling and the propagating density compressions associated with linear and NFM waves. The guided fast magnetosonic waves, together with density compressions and currents, are reproduced in the 3D pillar structure. We demonstrate or the first time the dynamic effects of the Lorentz force due to the magnetic shear in the non-force-free field on the pillar structure and on the propagation of the waves. The insights gained from the 3D MHD modeling are useful for improving coronal seismology of prominence structures that exhibit fast MHD wave activity.
\end{abstract}



\section{Introduction} \label{intro:sec}

Solar prominences \citep[also called filaments; e.g.][]{Tan95} are highly complex magnetic structures that extend from the photosphere up into the corona, where they support material that is much denser ($n\sim 10^{10-12}$ cm$^{-3}$) and cooler ($T\sim$ 1$\times$10$^4$ K) than the surrounding plasma ($n\sim 10^{8-9}$ cm$^{-3}$, $T\sim$ 1-2$\times$10$^6$ K).
High-resolution observations of prominences have been available in H I Balmer (H$\alpha$) and Ca II emission for decades from ground-based telescopes and, more recently, in various ion emission bands from satellite-borne instruments. These observations show that the prominence material is highly dynamic, exhibiting persistent flows, waves, and other oscillations, as well as MHD instabilities that can lead to its violent eruption \citep[for reviews, see][]{Lab11,Par14,Arr18}. Idealized models of quiescent prominences often assume an equilibrium magnetic structure that supports the cool material statically within the hot corona. The observations indicate that both the magnetic and thermal structures of prominences are often out of equilibrium, highly dynamic at small scales and gradually evolving at large scales.

Small-scale propagating and oscillating features in cool prominence threads and low-lying coronal loops have been studied from space for many years using high-resolution and high-cadence spectral observations. 
Hinode's Solar Optical Telescope \citep[Hinode/SOT;][]{Kos07} has observed such phenomena in H$\alpha$ and Ca~II emission \citep{Oka07,OW08,Sch13,Ofm15a,KOT18}, as has the Interface Region Imaging Spectrograph \citep[IRIS;][]{Dep14} in chromospheric Mg II emission as spectral lines and slit jaw images \citep{KOT18}. High-resolution prominence observations by Hinode/SOT  show that the prominences material exhibits constant down-flows, lateral flows, upflows, and dynamic evolution with the observed velocities in the range $1-100$ km s$^{-1}$ consistent with the effects of magneto-fluid instabilities \citep[e.g.,][]{Ber17}. Recent ground-based high-resolution observations using the New Vacuum Solar Telescope \citep[NVST;][]{Liu14} report evidence of small-scale oscillations and waves detected in H$\alpha$ in quiescent prominences \citep[e.g.,][]{Li18,Li22}. Advanced high-resolution resistive 3D MHD modeling of prominence structure evolution shows that the nonlinear development of the magnetic Rayleigh–Taylor instability produces small scale structures in the prominence material \citep{JK22}, and possibly can provide an alternative (to waves) interpretation of some of the observed small-scale oscillating structures.

The (quasi-) periodic, small-scale, oscillating features, with typical time scales of minutes in prominence threads and pillars, have been identified and modeled previously as linear fast magnetosonic waves \citep[e.g.,][]{Sch13}. Because the nonlinearity of these waves is evident in the observations in the form of steepening and asymmetric density compressions, the models were later extended to nonlinear fast magnetosonic (NFM) waves using an MHD model with two-dimensional spatial variations and three-dimensional vector fields \citep[2.5D MHD; see][]{Ofm15a,OK20}. The observed waves can be used to deduce the magnetic structure of prominences by applying techniques of coronal seismology \citep[e.g.,][]{NV05,Anf22}. These indirect methods are invaluable, as the coronal magnetic field is very difficult to measure directly using spectroscopic or other methods, while force-free extrapolation methods have limited applicability in realistic coronal structures. Coronal seismology remains based primarily on linear MHD wave theory. However, nonlinearity may significantly affect the wave structure, phase speed, wave dissipation, and couplings. Thus, interpreting observations of nonlinear waves requires the use of nonlinear wave theory or nonlinear MHD modeling for improved accuracy of the analysis.

Plasma flows, in addition to waves, are often observed in cool prominence threads in emission lines such as H$\alpha$ and Ca~II  \citep{Oka07,Ale13,Kuc14,Par14,Die18}. These flows may affect the oscillations through, for example, changes in the density that affect the phase speed of the waves and Doppler shifts of the oscillation frequencies. Recently, \citet{KOT18} and \citet{OK20} used Hinode/SOT Ca~II spectral lines to study small-scale motions in prominences. The observed propagating fluctuations were identified as NFM waves using a combination of data analysis and modeling. The observed NFM waves had typical periods $\sim$ 5-11 minutes and wavelengths $\sim$ 2000 km, while the flows had typical speeds $\sim$ 15-50 km s$^{-1}$.  The main properties of the observed NFM waves, combined with the effects of mass flows in prominence threads, were replicated by the model \citep{OK20}. The magnetic field strengths in the prominence were estimated to lie in the range 5-17 G. 

In the present study, we extend the previous studies of propagating waves in prominences with data analysis and modeling of a prominence pillar observed on 2012 February 14 from Hinode/SOT. We employ a new, fully three dimensional (3D) MHD model of small-scale NFM waves in an idealized prominence pillar with more realistic structure than in the previous studies. The new model allows us to investigate more complex fast magnetosonic wave generation, propagation, and interaction than in the previous 2.5D configurations, for example, by including the effects of magnetic shear,  and for the first time study the effects of non-force free shear magnetic field. The results are useful for interpreting high-resolution Hinode/SOT observations of prominence small-scale oscillations and for making further advancements in the coronal seismology of solar prominences using MHD waves.

Our paper is organized as follows. In \S\ref{obs:sec} we present new observations of propagating features in a prominence pillar. In \S\ref{model:sec} we describe the new 3D MHD model, along with the initial and boundary conditions used in the calculations. In \S\ref{num:sec}, we present the numerical results and compare them with observations. Finally, the discussion and our conclusions are given in \S\ref{dc:sec}.


\section{Observations and Data Analysis}
\label{obs:sec}

The studied prominence was observed by Hinode/SOT on 2014 February 14 from 10:48 - 13:15 UT as part of Hinode Operation Plan (HOP) 114. 
The data consisted of measurements from both the Broadband Filter Instrument (BFI) and the Narrowband Filter Instrument (NFI) \citep{Kos07,Tsu08}. The BFI was used to observe the \ion{Ca}{2} H line at 3969~\AA\, and the NFI was used to observe the \Ha\ line at 6563.2~\AA, both with cadences of 22.4~s.  The \Ha\ line positions for this data set were not well calibrated, but appear to be from near line center and in the blue wing of the line (about 416 m\AA\ from line center), making them not useful for Doppler measurements. 
The image field of view is about {112\arcsec} square, and the spatial resolution is 0.2-0.3\arcsec.
 Maps were processed with the  {\tt fg\_prep.pro} routine provided by to the  Solar Soft library (\url{https://www.lmsal.com/solarsoft/}, \citet{FH98}) by the Hinode team, including dark-current subtraction and flat-field removal. Drift and jitter were corrected using an image cross-correlation ({\tt fg\_rigidalign.pro}) routine.

For context, we inspected observations from the Global Oscillation Network Group to image the on-disk structure of the prominence in the days preceding its appearance at the limb. GONG \Ha\ images are provided by a network of six stations around the globe \citep{Har96} with a pixel size of about 1$\arcsec$.

\begin{figure}
\centerline{
\includegraphics[width=6cm, trim=20 50 0 0]{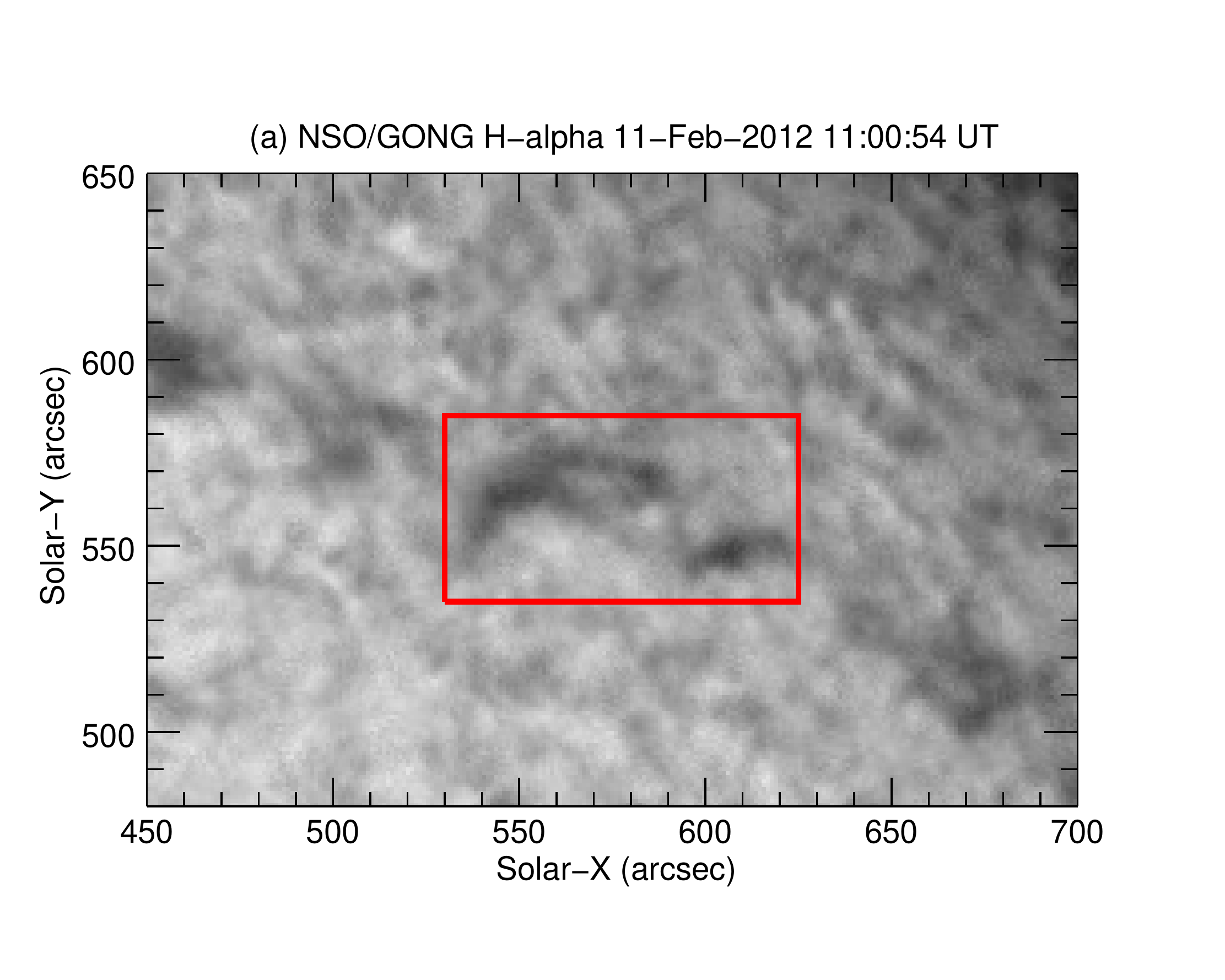}
\includegraphics[width=6cm, trim=60 50 40 0]{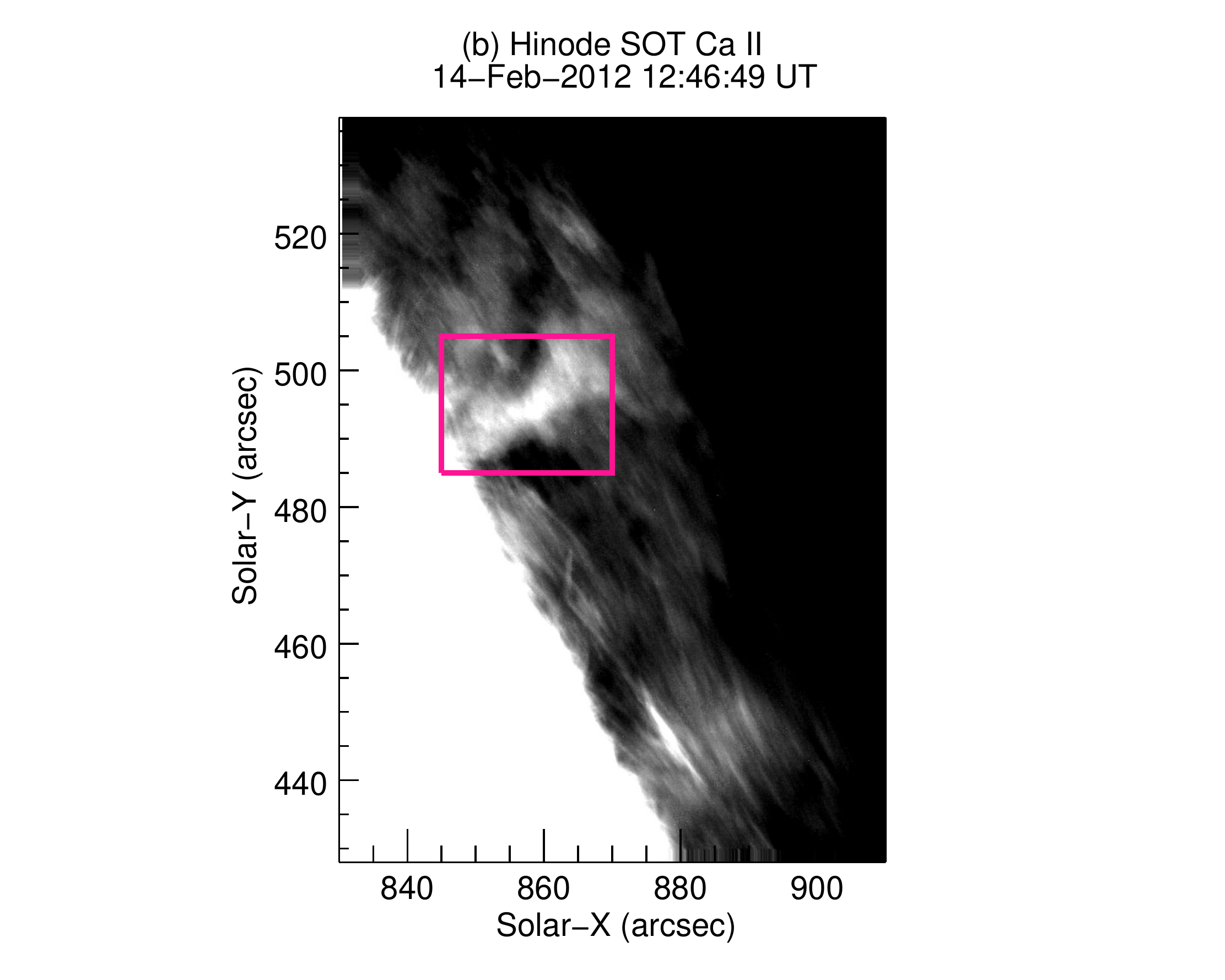}
\includegraphics[width=6cm, trim=60 50 40 0]{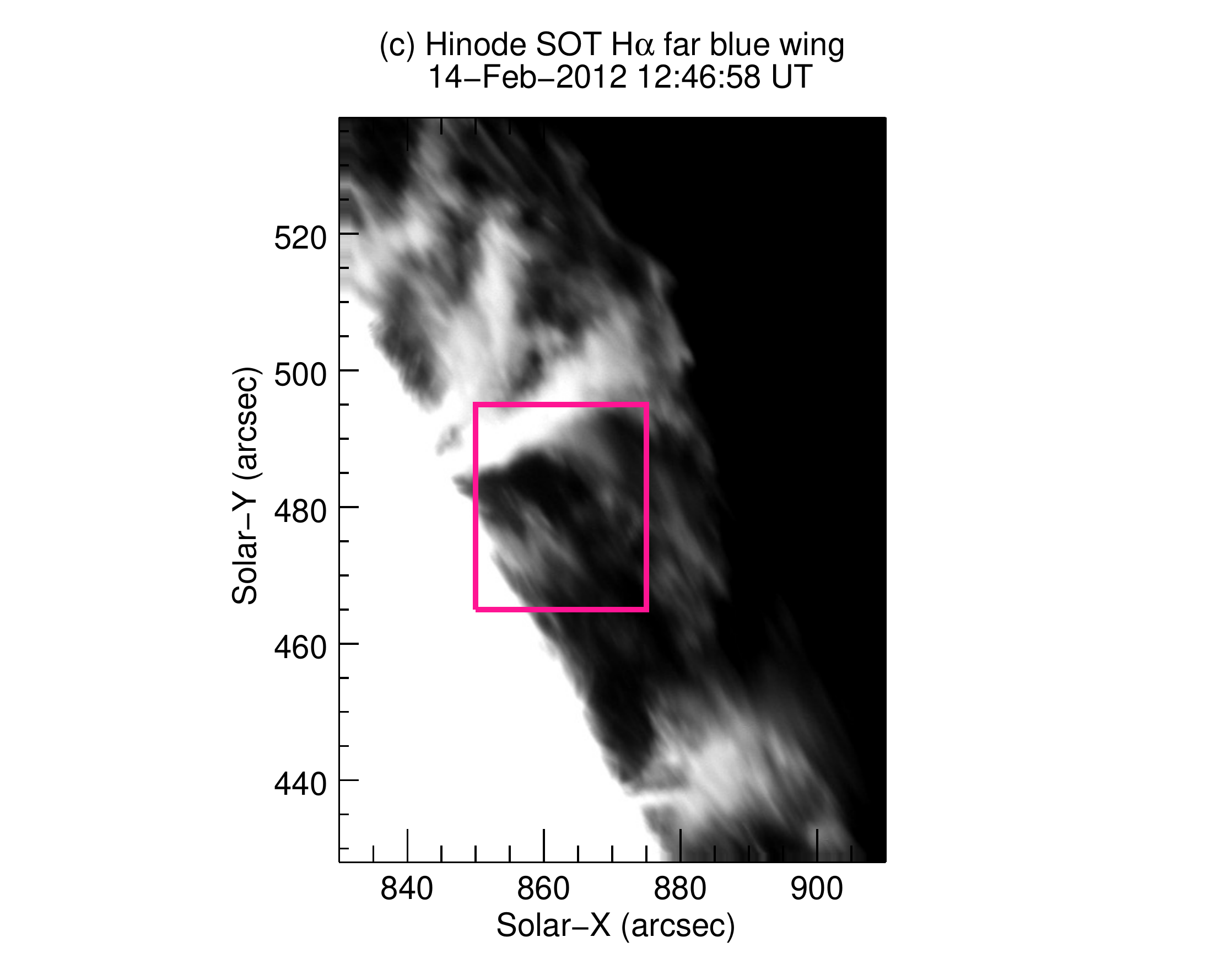}}
\caption{(a) GONG \Ha\ image showing the prominence on 11 Feb 2012, three days before the prominence was observed on the limb. The box shows the approximate field of view of the images in (b) and (c). (b) Hinode Ca~II image showing the prominence on the limb; the box is the field of view shown in Figure \ref{f:2012Feb14CaIIimg}. An animation that corresponds to this panel is available online.  The video shows the Hinode Ca~II emission observed on 14-Feb-2012 in the time interval 10:51:06-13:13:45 UT in an accelerated time  of 16 s. (c) H$\alpha$ far blue-wing image; the box is the field of view shown in Figure \ref{f:2012Feb14Haimg}.}
\label{f:ondisk_context}
\end{figure}

The features observed on the limb were part of a long prominence that extended more or less East-West above the northern active-region belt and curved equator-ward on the western end. A portion of that prominence seen against the solar disk three days before the observations we analyzed is shown in Figure~\ref{f:ondisk_context}a. The most evident prominence features seen in \Ha\ are a series of barbs connected by fainter spine flows. These barbs evolve over time, and it is difficult to identify individual barbs near the limb. However, the appearance of the region at the limb from Hinode/SOT, Figures~\ref{f:ondisk_context}b (Ca II) and \ref{f:ondisk_context}c (H$\alpha$), is consistent with associating the pillars with barbs that are oriented mostly along the line of sight. 

\begin{figure}
\includegraphics[width=6cm, trim=0 20 40 0]{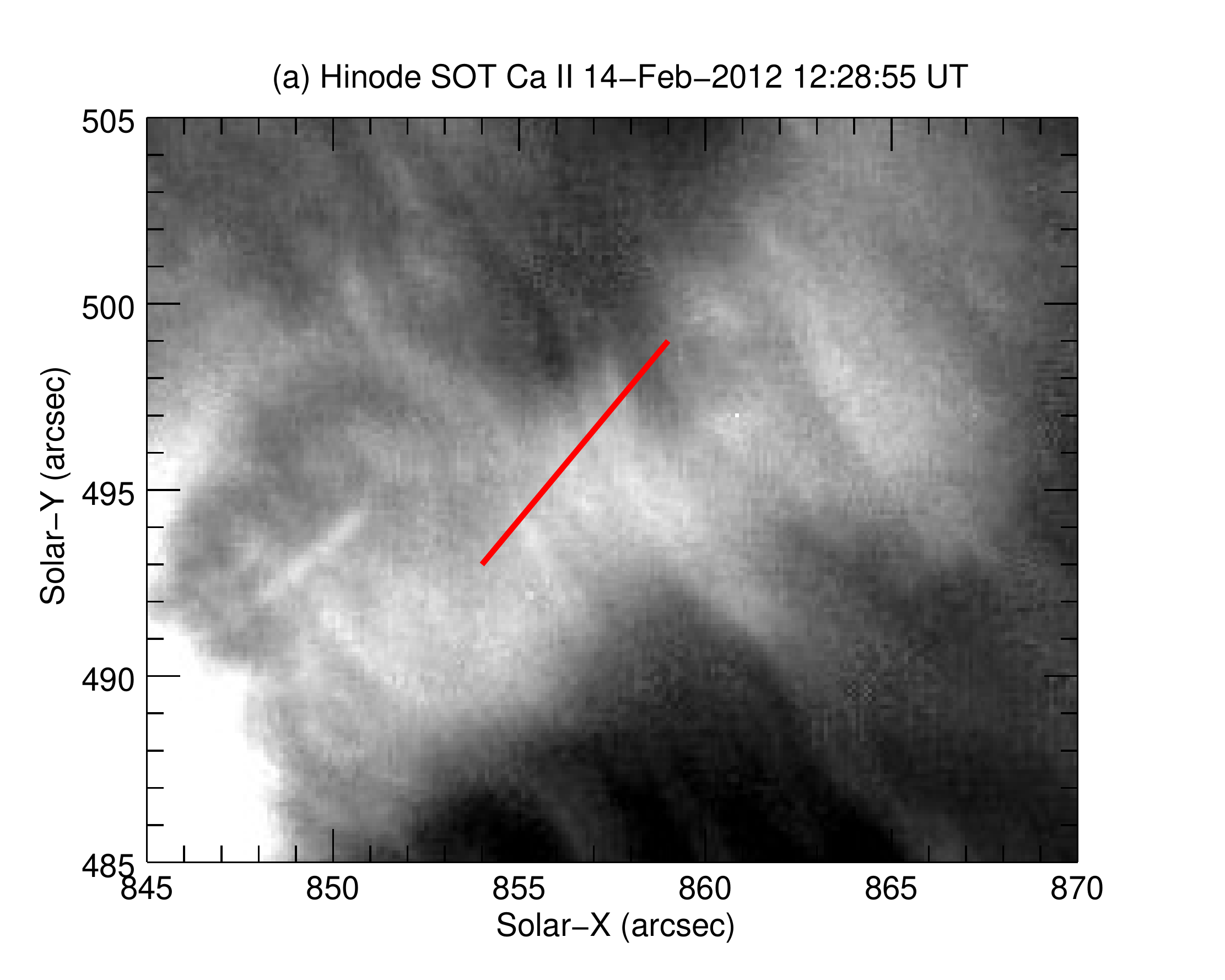}
\includegraphics[width=6cm, trim=0 40 0 0 ]{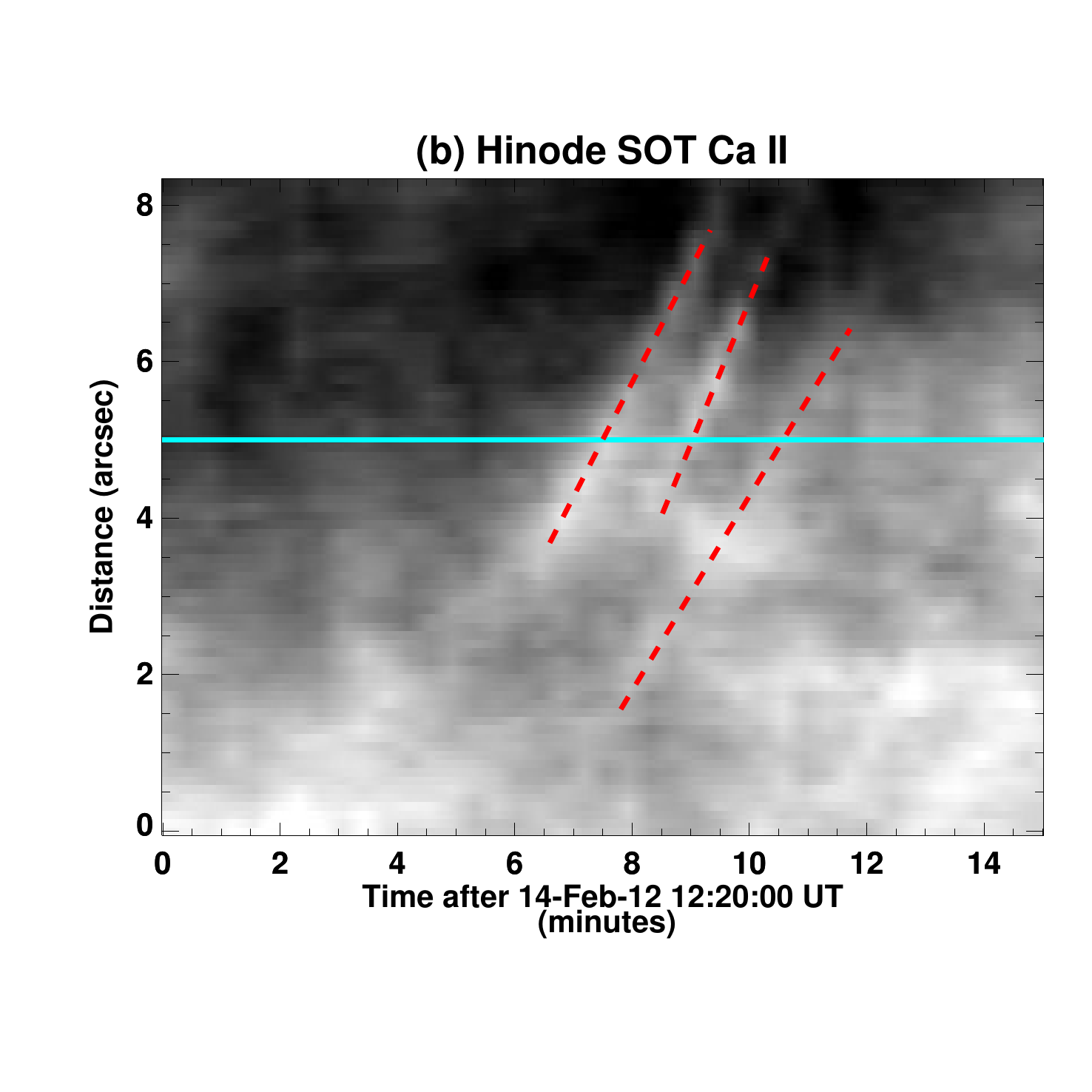}
\includegraphics[width=6cm, trim=0 0 0 0]{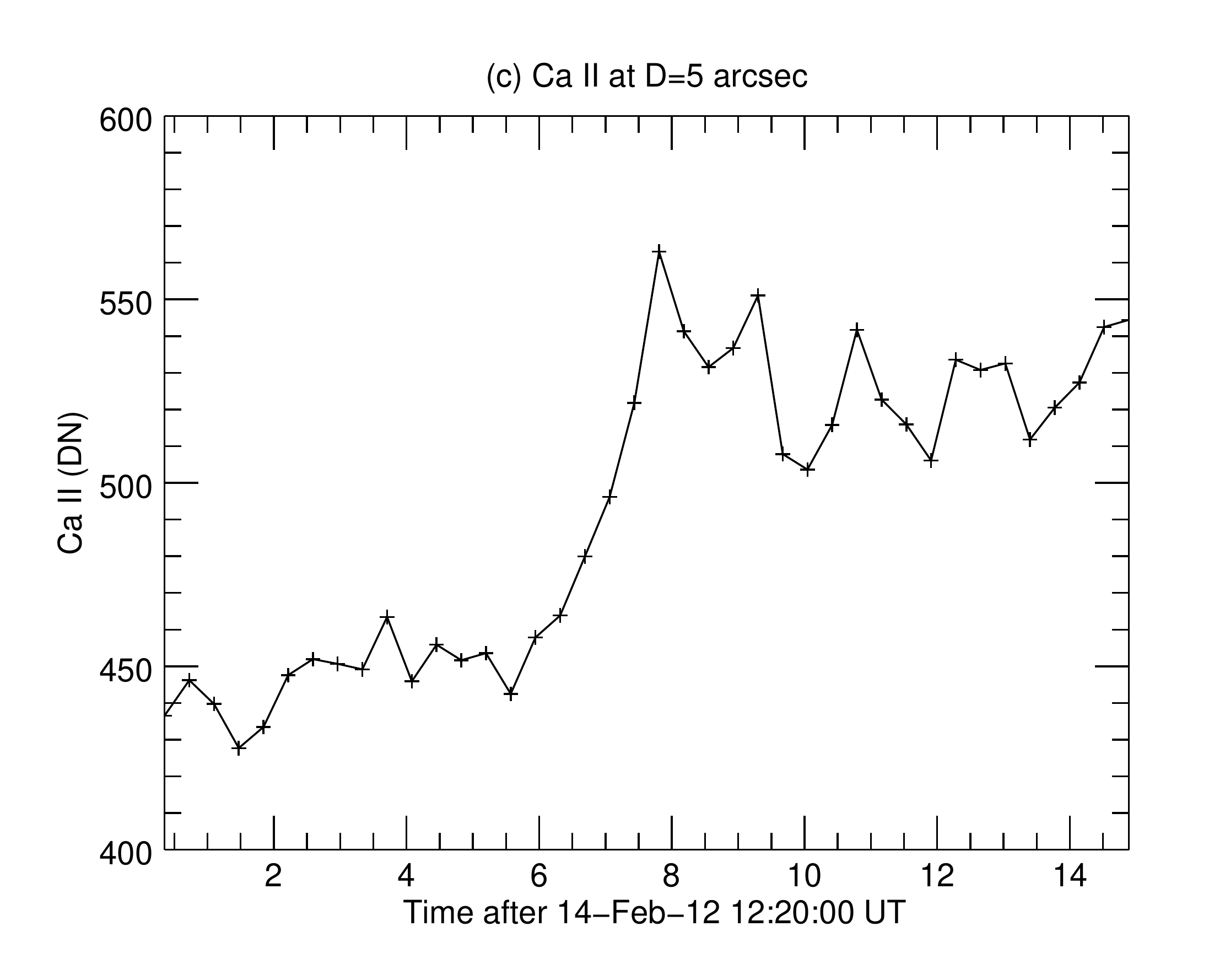}\\
\includegraphics[width=6cm, trim=0 20 50 0]{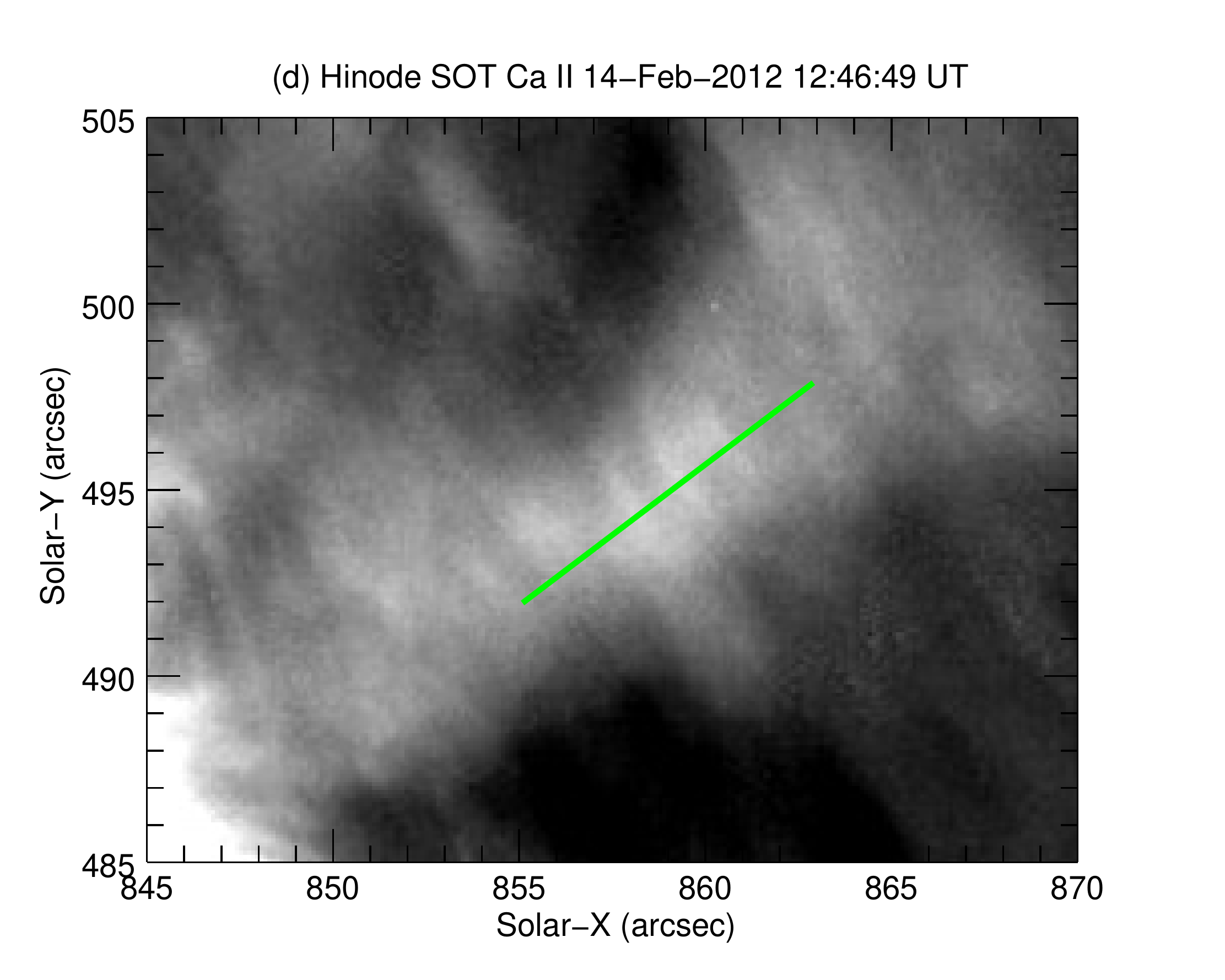}
\includegraphics[width=6cm, trim=0 40 0 0]{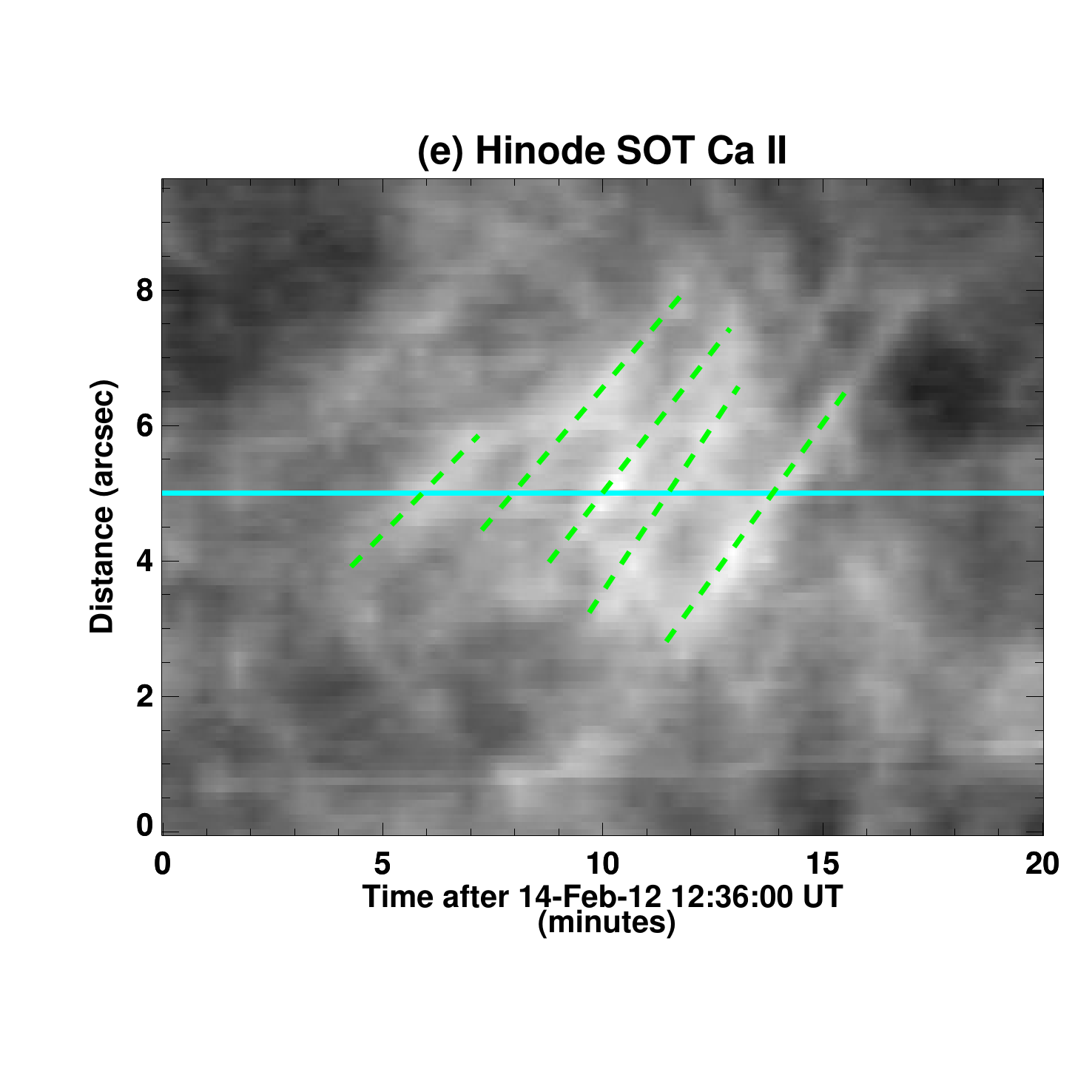}
\includegraphics[width=6cm, trim=0 0 0 0]{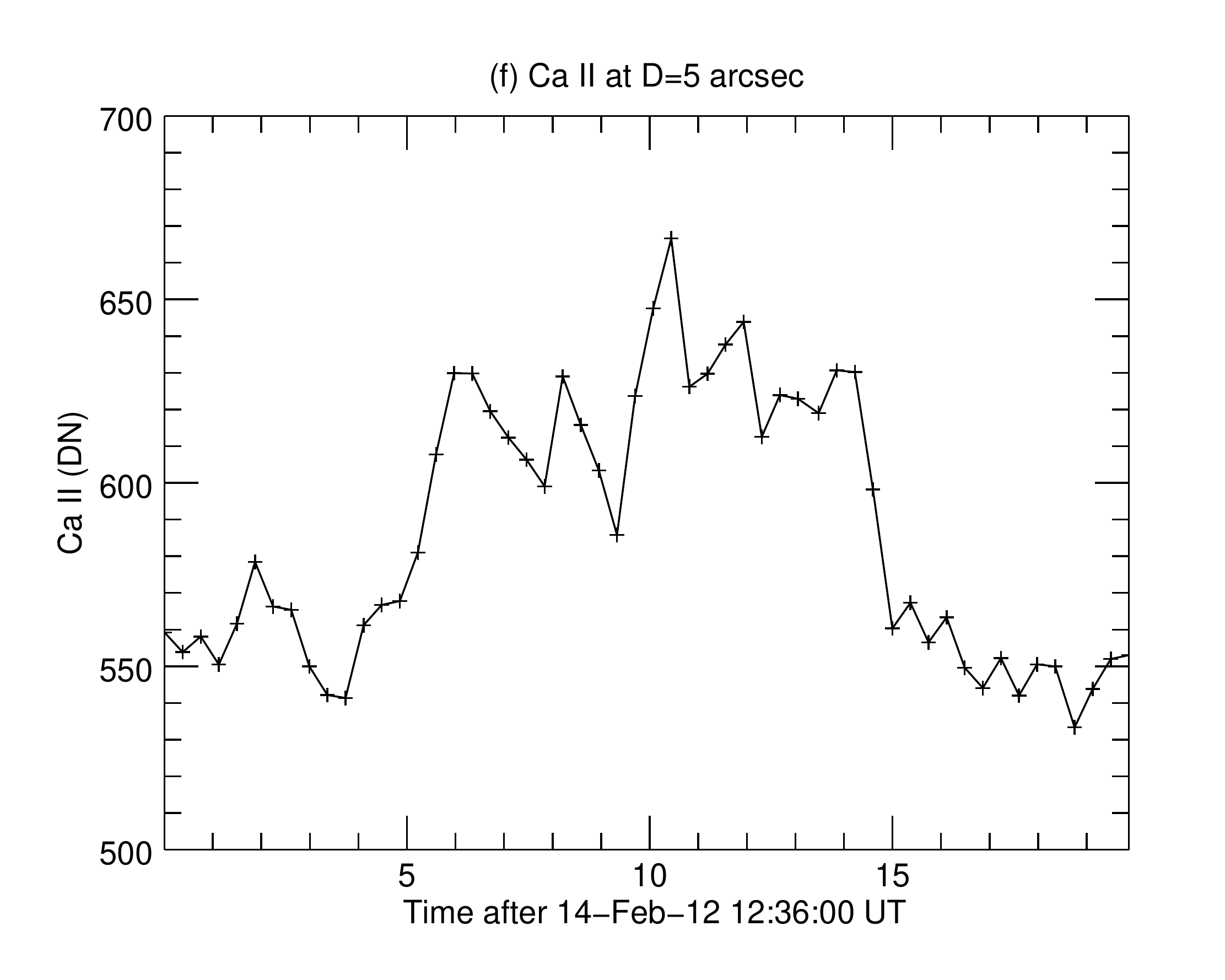}
\vspace{0.5cm}\caption{(a) Image of the Ca~II emission  obtained with Hinode/SOT on 14-Feb-2012 12:28:56UT. The solid red line indicates the data location for the time-distance diagram. (b) Time-distance diagram showing the propagation of the features along the solid red line in (a).  (c) Plots of intensity as a function of time at the locations shown with the blue horizontal line in (b). (d)-(f) The same for a different set of pulses obtained along the solid green line in (d). The slopes of the dashed red (b) and green (e) lines indicate the propagation speed of the pulses in the plane of the sky. A video showing the field of view in panels (a) and (d) is included online.  The video shows the Hinode SOT Ca II intensity observed on 14-Feb-2012 in the time interval 12:18:06 UT to 15:56:59 UT in an accelerated time of 4 s.}
\label{f:2012Feb14CaIIimg}
\end{figure}

Figure \ref{f:2012Feb14CaIIimg} shows time-distance diagrams for the small-scale propagating features (i.e., `pulses')  observed in the Ca~II images in two locations. The pulses were measured along a 5-pixel-wide area centered on the solid red and green lines shown in panels (a) and (d), respectively.
Panel (b) shows a series of pulses with plane-of-sky velocities 12-16 km s$^{-1}$ determined from the slopes of the dashed red lines, which were visually fit to the intensity peaks. The peaks are about 1 min apart, and the distances between pulses are in the range 1330-2030 km.
Panel (e) shows another set of pulses corresponding to the location shown in panel (d). These pulses have peaks 1.5-2.3 min apart, velocities 8-11 km s$^{-1}$ obtained from the slopes of the dashed green lines, and distance between pulses in the range 800-1600 km. Panels (c) and (f) show the intensities along the horizontal lines shown in panels (b) and (e) respectively. The variations between the maximum and minimum intensities of the individual features are about 10\% of the total intensity.

\begin{figure}
\hspace{-0.6cm}\includegraphics[width=6.5cm, trim=40 20 40 0]{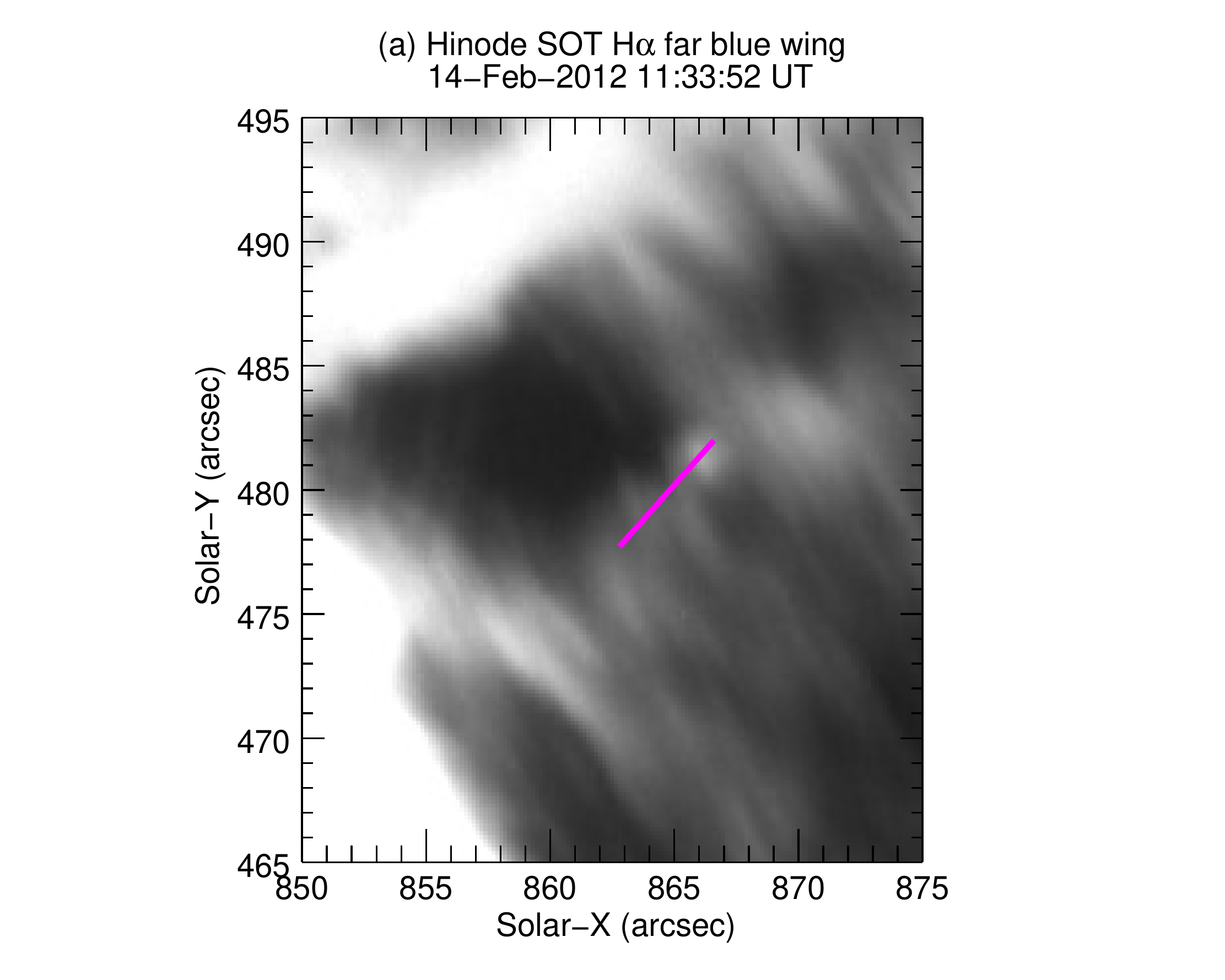}
\hspace{-0.4cm}\includegraphics[width=6cm, trim=50 50 10 0]{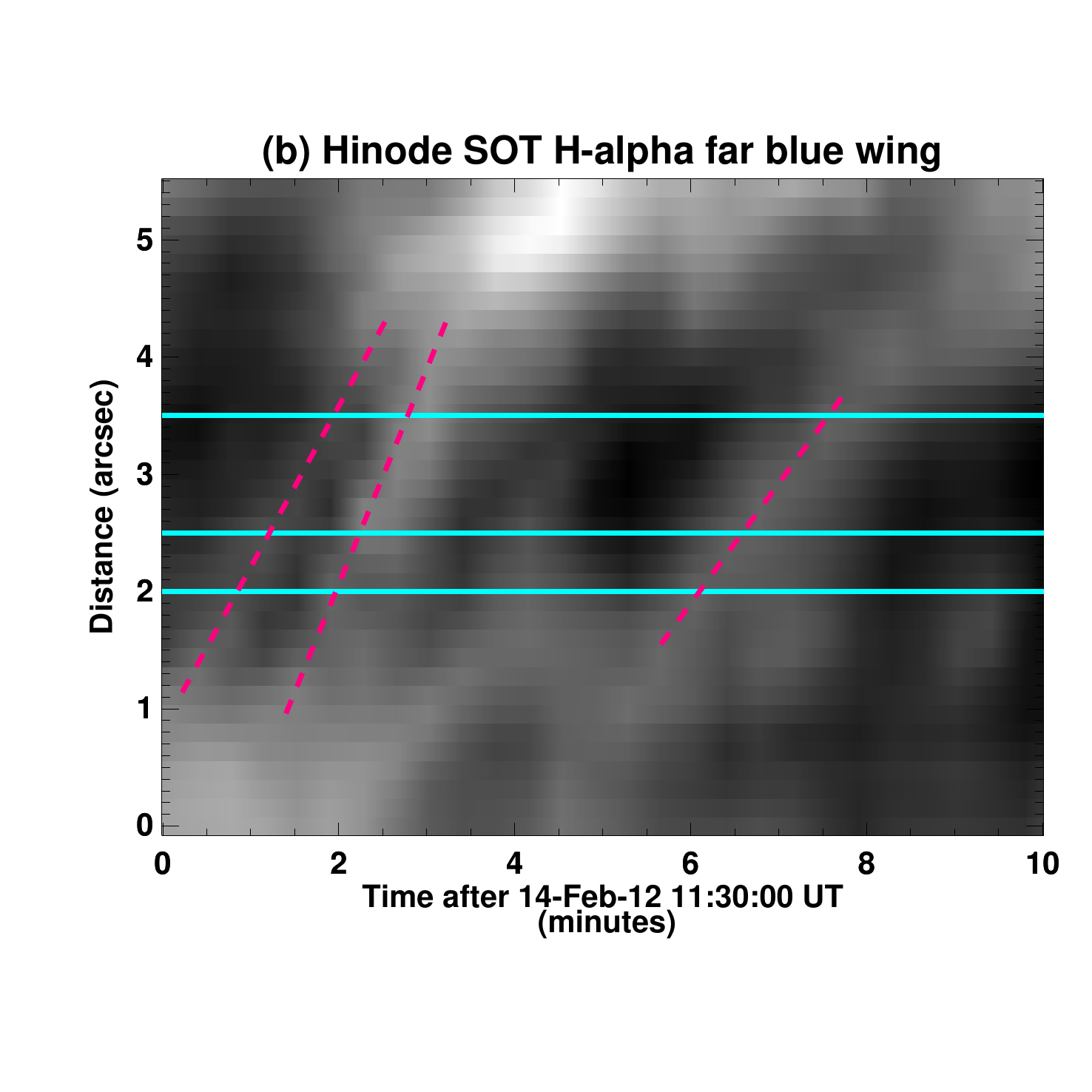}
\hspace{0.3cm}\includegraphics[width=5.75cm, trim=0 0 0 0]{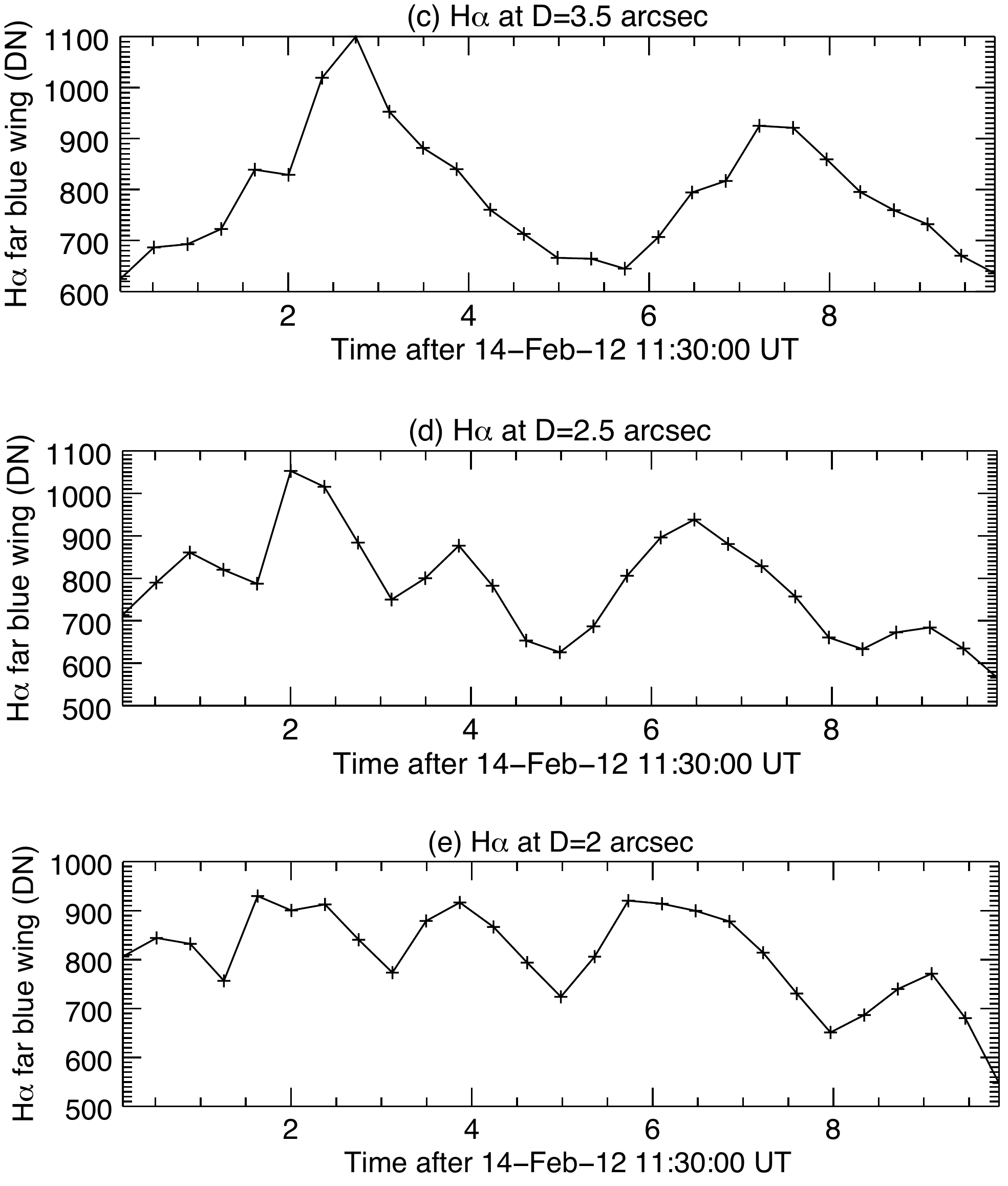}

\caption{(a) Image in the far blue wing of \Ha\ obtained with Hinode/SOT on 14-Feb-2012 at 11:33:52 UT. The solid purple line shows the location of the data for the time-distance diagram. (b) Time-distance diagram showing the propagation of the pulses along the solid purple line indicated in (a). (c)-(d) plots of intensity as a function of time at the locations shown with the blue horizontal lines in (b). The slopes of the dashed purple lines in (b) indicate the propagation speed of the pulses in the plane of the sky.  A video that corresponds to the field of view in panel (a) is included online. The observed Hinode SOT \Ha\ far blue wing field of view on 14-Feb-2021 in the time interval 11:25:17-11:42:26UT is shown in an accelerated time of 2 s.
}
\label{f:2012Feb14Haimg}
\end{figure}

Figure \ref{f:2012Feb14Haimg} shows time-distance diagrams for moving features seen in the \Ha\ blue wing. Shown are a series of pulses with plane-of-sky velocities 12-16 km s$^{-1}$, peaks 1-5 min apart with sharp non-sinusoidal peaks indicative of nonlinear steepening, and distances between pulses of 1000-3000 km. The plane-of-the-sky propagation speed is likely reduced compared to the `true' phase speed due to projection effects, and the value is in qualitative agreement with possible fast magnetosonic speeds in cool prominence material of the order $\sim$ 20 km s$^{-1}$ \citep[see, e.g.,][]{Sch13}. The variations between the maximum and minimum intensities of the individual features are about 30-60\% of the total intensity.

\begin{figure}[ht]
\centerline{\includegraphics[height=9in]{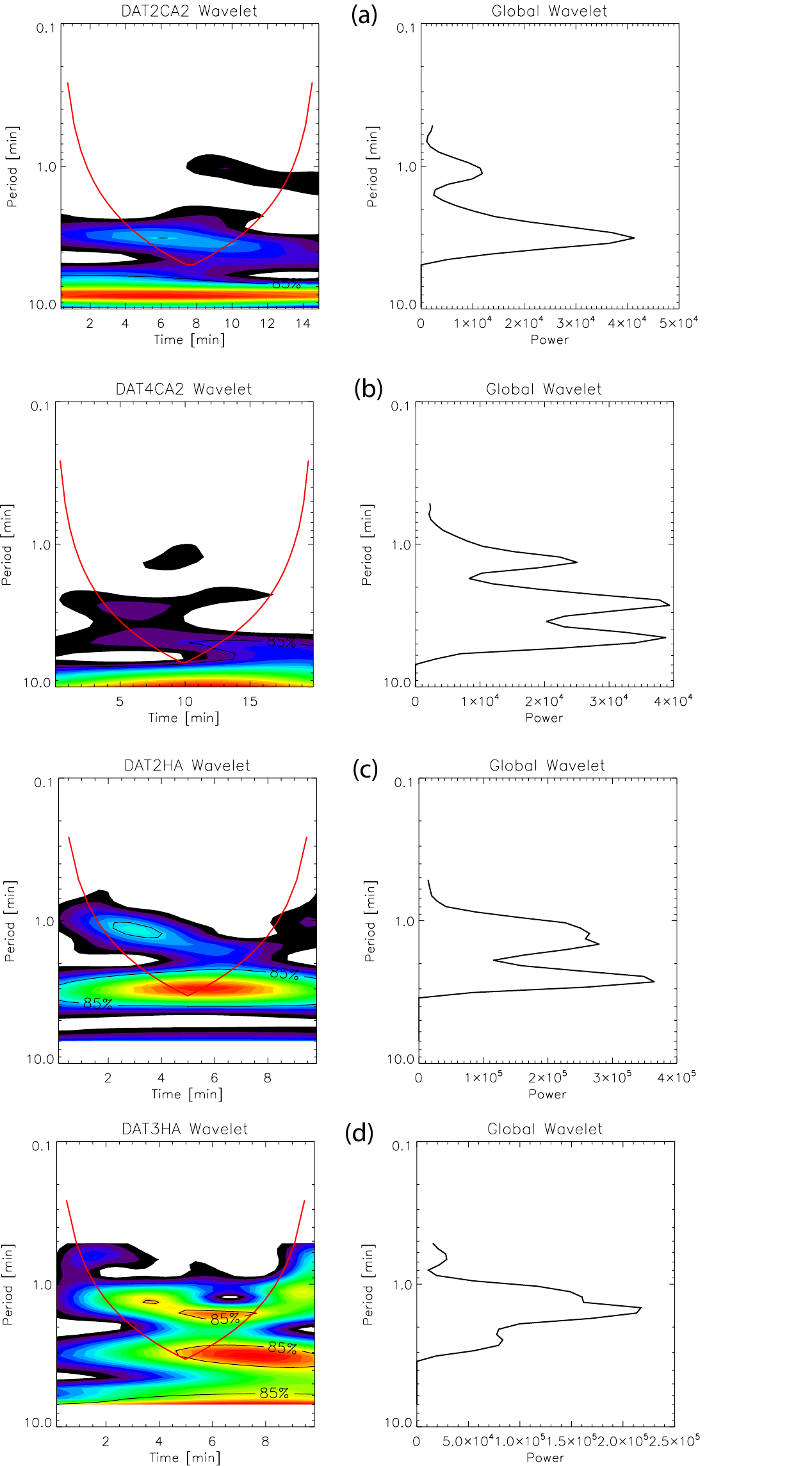}
}
\caption{The results of the wavelet analysis of the oscillations shown in Figures (a) \ref{f:2012Feb14CaIIimg}c; (b) \ref{f:2012Feb14CaIIimg}f (c) \ref{f:2012Feb14Haimg}d; (d) \ref{f:2012Feb14Haimg}e;  respectively. The Morlet wavelet was used, and the 85\% confidence level contour is indicated on the wavelet power. The cones of influence where boundary effects may affect the results are indicated with the red curve on each wavelet panel. The global wavelets in the cone of influence for each case are shown in the corresponding right panels.} 

\label{wavelet:fig}
\end{figure}

We have performed a wavelet analysis \citep{TC98} of the oscillations in Ca~II and the far blue wing of \Ha\ cuts shown in Figures~\ref{f:2012Feb14CaIIimg} and \ref{f:2012Feb14Haimg} using the Morlet wavelet. In Figure~\ref{wavelet:fig} we show the results of the analysis with evident highest confidence level for the wavelet magnitude greater than $85\%$. The cones of influence indicate the regions that may be affected by the boundaries. The results show the global wavelet power integrated inside the cone of influence, indicating significant power in $\sim 1-3$ min period oscillations, consistent with the temporal evolution at the indicated temporal cuts, and in the animations of the observed oscillations included online. The wavelet analysis and the global wavelets provide unbiased quantification of the observed oscillations and their statistical significance.

Thus, we observe multiple cases of a short series of oscillatory features propagating in a direction roughly away from the limb in the plane of the sky, separated by $\sim$ 1 min. Each individual feature is slightly elongated perpendicular to the direction of motion, hence is similar to features described previously by others \citep{Sch13,Ofm15a,KOT18}. Because the pillars are likely to be elongated structures along the line of sight, these moving features may be related to motions observed in different (perpendicular) line of sight in extended prominence structures as transverse oscillations combined with flows of cool material \citep{OW08,Oka16,OK20}. 

\section{Numerical 3D MHD Model, Boundary Conditions, and Parameters} \label{model:sec}

In order to model the NFM waves in a prominence pillar we solve the resistive 3D MHD equations using our code NLRAT described in detail in previous papers \citep{OT02,POW18,OL18,OW22}. The normalized resistive MHD equations with gravity, using standard notation for the variables, are
\begin{linenomath*}
\begin{eqnarray}
&&\frac{\partial\rho}{\partial t}+\nabla\cdot(\rho\mbox{\bf V})=0,\label{cont:eq}\\
&&\frac{\partial(\rho\mbox{\bf V})}{\partial t}+\nabla\cdot\left[\rho\mbox{\bf V}\mbox{\bf V}+\left(E_up+\frac{\mbox{\bf B}\cdot \mbox{\bf B}}{2}\right)\mbox{\bf I}-\mbox{\bf BB}\right]=-\frac{1}{F_r}\rho\mbox{\bf F}_g,\label{mom:eq}\\
&&\frac{\partial\mbox{\bf B}}{\partial t}-\nabla\times(\mbox{\bf V}\times\mbox{\bf B})=\frac{1}{S}\nabla^2\mbox{\bf B},\label{ind:eq}\\
&&\frac{\partial(\rho E)}{\partial t}+\nabla\cdot\left[\mbox{\bf V}\left(\rho E+E_up+\frac{\mbox{\bf B}\cdot \mbox{\bf B}}{2}\right)-\mbox{\bf B}(\mbox{\bf B}\cdot\mbox{\bf V})+\frac{1}{S}\left(\nabla\times\mbox{\bf B}\right)\times\mbox{\bf B}\right] =-\frac{1}{F_r}\rho\mbox{\bf F}_g\cdot\mbox{\bf V}.\label{ener:eq} 
\end{eqnarray}
\end{linenomath*}
With our normalization $E_u = \beta/2$ is the magnetic Euler number (ratio of thermal pressure to Alfv\'en-wave pressure), $F_r=V_A^2R_s/(GM_s)$ is the magnetic Froude number (ratio of magnetic force to gravitational force), where $G$ is the gravitational constant,
$M_s$ is the solar mass, $R_s$ is the solar radius, and $S$ is the Lundquist number (ratio of resistive diffusion time to Alfv\'en time). The details of the normalization of the variables can be found in \citet{OL18}. The gravitational force, 
\begin{eqnarray}
\mbox{\bf F}_g=\frac{a_0^2}{(R_s+z-z_{min})^2}\mbox{\bf\^z},\label{grav:eq}
\end{eqnarray}
is modeled with the assumption of small height of the prominence compared to the solar radius $R_s$, where $a_0=0.1R_s$ about 70 Mm is the normalization length scale of the coordinates, and $z_{min}$ is the height of the lower boundary in the model.  We note that in the present model we have excluded radiative losses and thermal conduction, and the prominence pillar structure is provided as an initial state, rather than produced self-consistently by the model. The total energy density is given by 
\begin{eqnarray}
\rho E=\frac{E_up}{(\gamma-1)}+\frac{\rho V^2}{2}+\frac{B^2}{2}. 
\end{eqnarray}
In the present model, we neglect radiative cooling and thermal conduction because these losses are small on the typical time scales of the NFM waves. 
For coronal temperature $T=1\times10^6$ K, density $n=10^9$ cm$^{-3}$, and magnetic field magnitude $B=10$ G we obtain the Alfv\'{e}n speed $V_A=690$ km s$^{-1}$, the Alfv\'{e}n time $\tau_A=101$ s, the plasma $\beta \approx 0.07$, the Froude number $F_r=0.25$, and the Euler number $E_u=3.47\times10^{-2}$ (for the case with $B_0=20$ G, $E_u$ is reduced by a factor of four, to  $E_u=8.67\times10^{-3}$). Note that, for uniform magnetic field, the value of $\beta$ is identical to the coronal value all across the prominence pillar, due to the uniform thermal pressure along the magnetic field lines that cross the pillar. For computational stability purposes, the effect of gravity in the model is reduced by a factor of 10 by correspondingly increasing $F_r$, in order to slow the gravitational settling of the cool material in the prominence pillar. The reduced gravity does not affect the results significantly, since the dominant restoring force of  the oscillations is the Lorentz force (i.e., magnetic field-line `tension').  In the above equations we have neglected viscosity, radiative losses, and thermal conduction. The resistive terms are used with the Lundquist number set to $S=10^5$, which does not affect the results significantly on the NFM time scales. An empirical value of the nearly isothermal polytropic index, $\gamma=1.05$, is used that accounts for coronal heating. These modeling parameters improve the stability of the background prominence pillar structure on the time scale of MHD wave propagation, without affecting significantly the NFM wave dynamics. 

\begin{figure}[ht]
\centerline{
\includegraphics[width=0.5\linewidth]{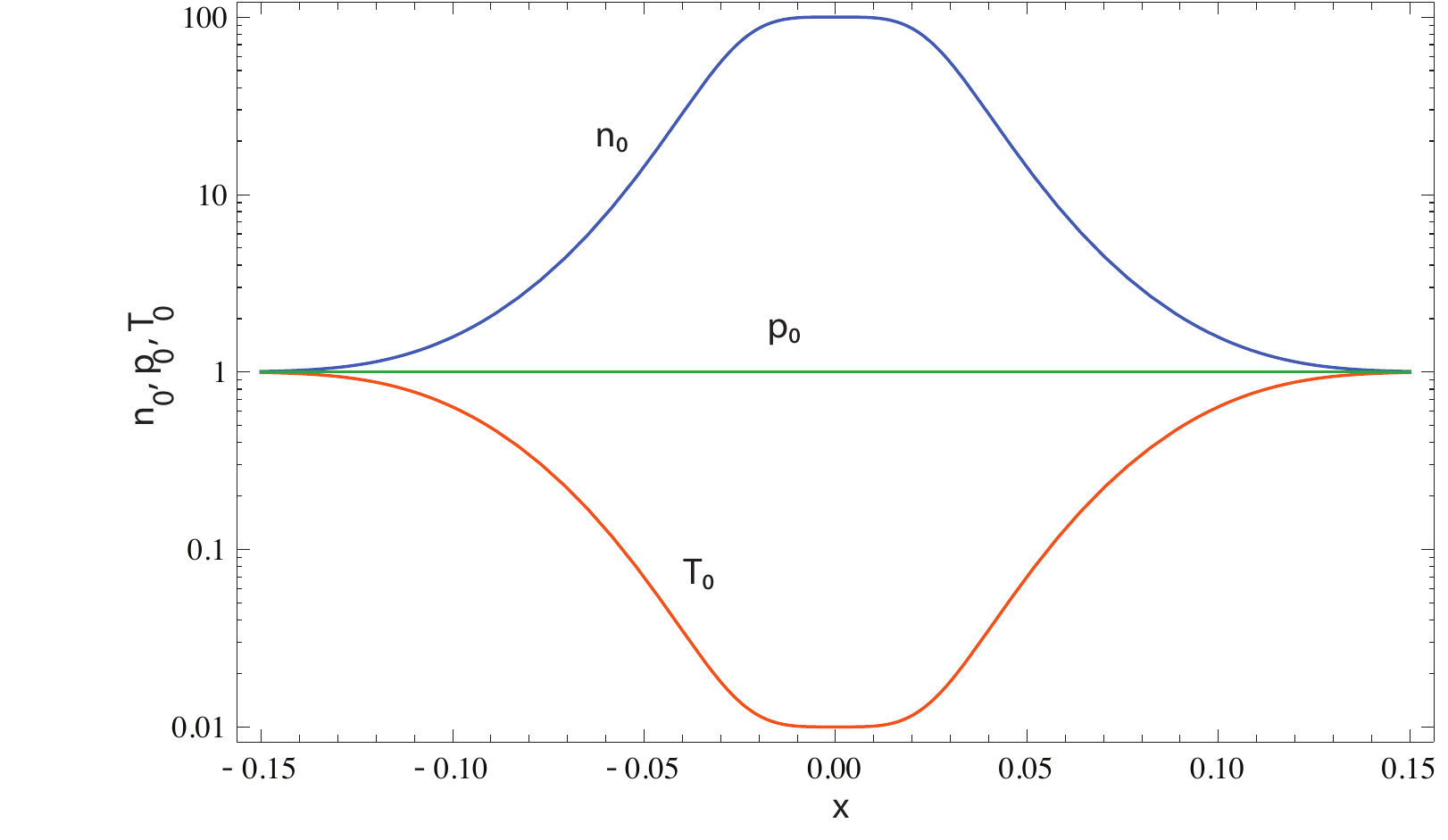}
}
\caption{The $x$ dependence (across the prominence pillar) of normalized initial temperature, $T_0$ (red), density, $n_0$ (blue), and thermal pressure $p_0$ (green) in the model prominence and surrounding corona. {The magnitudes of the variables are shown on $Log$ scale.}}
\label{n0_T0:fig}
\end{figure} 

The initial $x$-dependent temperature $T_0$ and density $n_0$ structures are given by 
\begin{eqnarray}
&&T_0(x)=T_{max}-(T_{max}-T_{min})e^{-[(x-x_0)/w]^{2q}},\label{T0:eq}\\
&&n_0(x)=p_0/T(x),\label{n0:eq} 
\end{eqnarray}
where the coronal temperature is $T_{max}$, the prominence temperature is $T_{min}$, the exponent $q=2$ defines the sharpness of the temperature transition between the corona and the prominence pillar, $w=0.05$ is the half-width of the prominence pillar, and $x_0=0$ is the center position of the pillar. The initial temperature and density dependencies on the $x$ coordinate across the model prominence pillar are shown in Figure~\ref{n0_T0:fig}. The normalized thermal pressure, $p_0=n_0T_0=1$, is uniform. In the present model we use $T_{min}/T_{max}=0.01$, consistent with the typical ratios of the prominence to corona temperatures. It is evident in Figure~\ref{n0_T0:fig} that $T_0$ decreases from its coronal value by  two orders of magnitude, while $n_0$ increases correspondingly by the two orders of magnitude in the prominence pillar. Note, that the fine-scale structuring of the background density in the direction of wave propagation, i.e., with height, may introduce dispersion, enhanced damping, and small deceleration of the fast magnetosonic waves \citep[e.g.,][]{Mur01}. In the present model the fast magnetosonic speed is lower by a factor of 10 in the pillar compared to the surrounding corona, and the speed could be even lower in higher density cool prominence structures. The prominence-corona transition region (PCTR) \citep[see the review,][]{Par14} along the magnetic field is evident in the model, with the length computed as the difference between the half width at half maximum (HWHM) of the prominence pillar and the half width at 10\%  of peak density as $\sim 1$ Mm in physical units. This value of PCTR thickness is consistent with previous studies \citep[e.g.,][]{Chi92,Gun11}. In normalized units the mass density is equal to the number density $\rho_0=n_0$. The initial state is in equilibrium when the magnetic field is uniform in the $x$ direction without gravity, which was first used to study prominence oscillations by \citet{JR92} and later in 2.5D MHD models of NFM waves in prominences \citep{Ofm15a,OK20}. Here we adopt this initial state in the 3D MHD model, as well as study additional magnetic configuration that depart from equilibrium. Since we consider the effects of reduced solar gravity the initial state is not strictly in equilibrium. The initial nonequilibrium leads to formation of gradients in the initially uniform magnetic field that produce a Lorentz force balancing gravity. However, the departure from equilibrium is small in the low-$\beta$ prominence model, as shown below (see, Case~0 in Section~\ref{num:sec}). While the initial state of the density is uniform in the $y$ and $z$ directions, transverse variability is introduced by the effects of the source of the waves (i.e., boundary conditions), in addition to the effects of gravity. Since the source of the waves at the lower boundary of the prominence pillar depends on $x$ and $y$ and on time (see, Equation~\ref{Vzt0:eq} below), the compressional fast magnetosonic wave pressure introduces structure primarily in the density and magnetic field in $x$, $y$, and $z$ directions inside the pillar.

\begin{figure}[ht]
\includegraphics[width=0.5\linewidth]{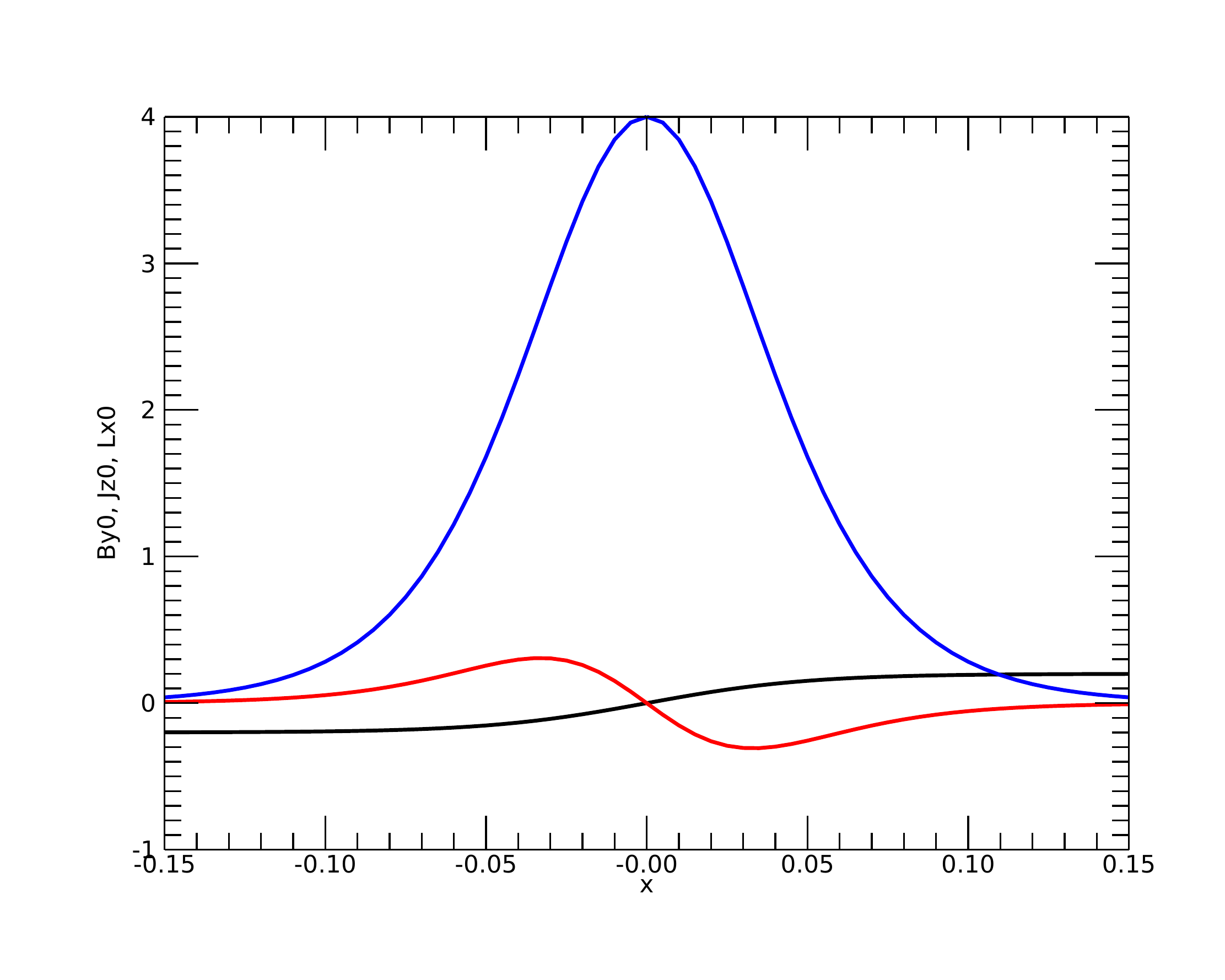}
\includegraphics[width=0.5\linewidth, ,trim={0 2cm 0 0},clip]{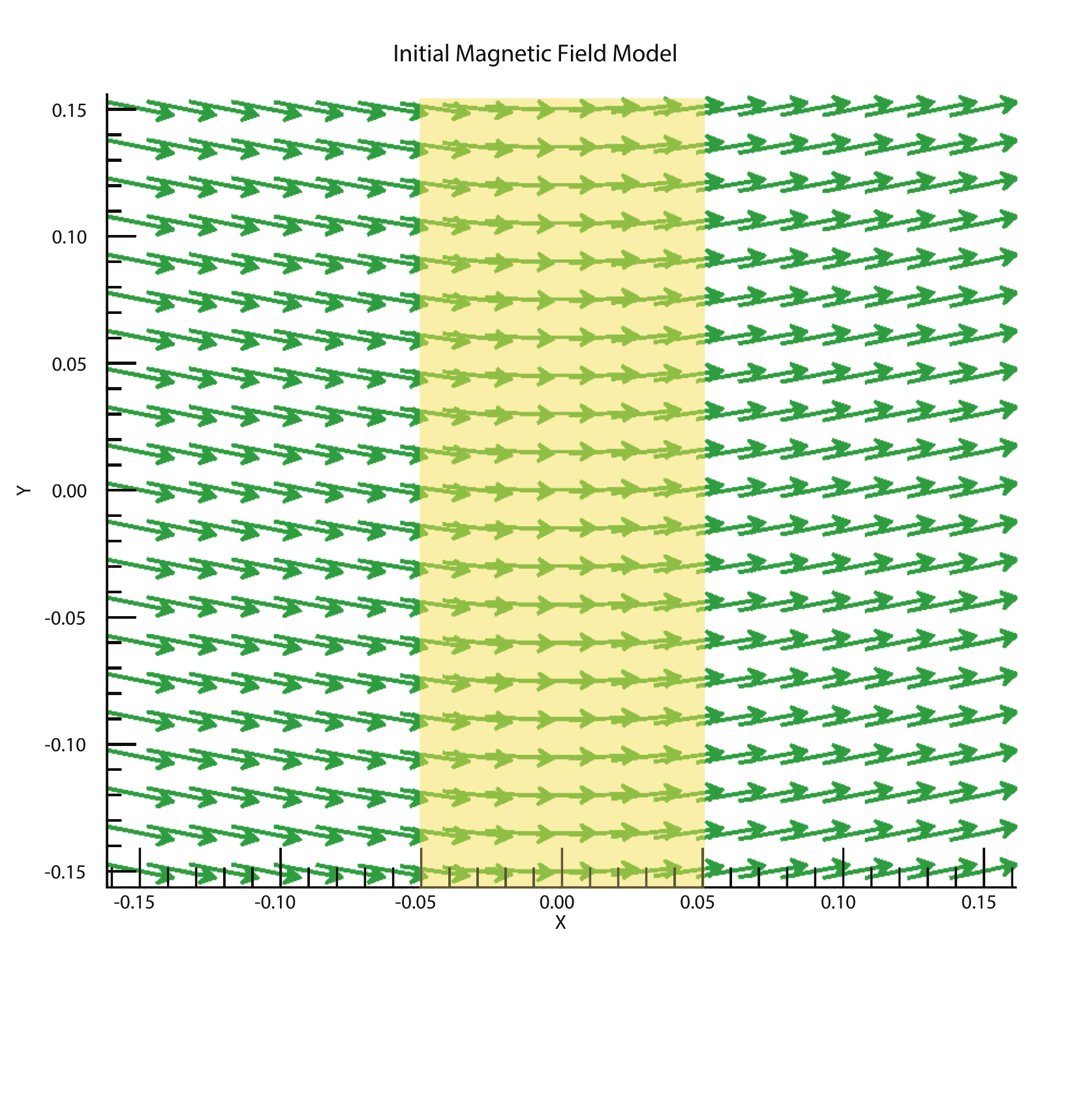}

\caption{The initial magnetic field with $B_{y0}=0.2$ (see Case~7, below).  (a) The $x$ dependencies of the $y$ component of the magnetic field $\mbox{\bf B}_{0}$ (black), the corresponding $z$ component of the current density $\mbox{\bf j}_{0}$ (blue), and the resultant $x$ component of the Lorentz force $\mbox{\bf L}_{0}$ (red). (b) The initial magnetic field vectors in the $x$-$y$ plane. The shaded area indicates schematically the location of the prominence pillar.}
\label{B0_xy:fig}
\end{figure}

 Realistic three-dimensional force-free extrapolations show that
magnetic field of dips in quiescent prominences is mostly horizontal \citep[e.g.,][]{Dud12}. Observed prominence structure shows evidence of magnetic shear and flows (see, e.g., \citet{Ant94} and the recent review by \citet{Gib18}). Our aim is to investigate the effects of uniform as well as sheared  magnetic field on the propagation of nonlinear fast magnetosonic waves in the prominence pillar. There are many past observations of flows in prominence pillars \citep[e.g.,][and references within]{OK20}. While there could be several possible sources for the observed flows in prominences of jet-like or large-scale flows, here, we model the effects of an unbalanced Lorentz force (i.e., non-force free magnetic configuration) with small shear as the driver of the large-scale flows in the prominence foot, Cases~4-8. While in some observations of Polarity Inversion Lines (PIL) in prominences the magnetic shear could be large and the magnetic field possibly force-free, modeled with linear force-free field magnetic field \citep[e.g.,][]{AD98}, or nonlinear force-free field \citep[e.g.,][]{Jia14}, our model investigates for the first time the effect of non-force-free field on the formation of large-scale flows and on the propagation of fast magnetosonic waves in the prominence pillar self-consistently. Our model reproduces the main properties of such sheared magnetic configurations by introducing the $x$-dependent $B_y$ component that changes sign in the center of the prominence pillar at x=0, as modeled by Equation~\ref{B0:eq}, 
\begin{eqnarray}
&&\mbox{\bf B}_0=B_{x0}\mbox{\bf\^x}+B_{y0}{ \rm tanh}(x/w)\mbox{\bf\^y},
\label{B0:eq}
\end{eqnarray}
where $B_{x0}$ and $B_{y0}$ are given in Table~\ref{param:tab} for the eight cases studied, and $w=0.05$ is the fixed half-width of the prominence pillar. When $B_{y0}=0$, the magnetic field is potential, and the initial state given by Equations~\ref{T0:eq} and \ref{n0:eq} is in equilibrium. In order to study initial states that depart from equilibrium and contain currents (non-force-free), we use small values of $B_{y0} \ll B_{x0}$ in the initial state. The corresponding current density $\mbox{\bf j}_0$ and Lorentz force $\mbox{\bf L}_0$ in the $x$-$y$ plane are given by 
\begin{eqnarray}
&&\mbox{\bf j}_0=\nabla\times\mbox{\bf B}_0=\frac{B_{y0}} {w}\mbox{\rm sech}^2(x/w)\mbox{\bf\^z},\\
&&\mbox{\bf L}_0=\mbox{\bf j}_0\times\mbox{\bf B}_0=j_{z0}(-B_{y0}\mbox{ \rm tanh}(x/w)\mbox{\bf\^x}+B_{x0}\mbox{\bf\^y}).
\label{j0L0:eq}
\end{eqnarray}
The $x$ dependence of $\mbox{\bf j}_0$ and $\mbox{\bf L}_0$ along with the corresponding magnetic field in the $x$-$y$ plane are shown in Figure~\ref{B0_xy:fig}.  The present model produced the desired Lorentz force that points toward the center of the pillar. Note, that we have also experimented with other forms of $B_y$, such as a centrally peaked profiles, and found similar results for the fast magnetosonic waves, but with different forms of the Lorentz force and directions of the large-scale flows. The location of the prominence pillar is depicted by the yellow-shaded area.

The adopted form of $B_y$ is justified by the dynamics of the observed flows and allows exploring the fast magnetosonic wave propagation effects in non-pontential and non-force-free magnetic field in a prominence. Moreover, this magnetic configuration may correspond qualitatively to a section of a sigmoidal filament structure that is often unstable, leading to eruption \citep[e.g.,][]{Yan12,Dau21}. 

The vertical extent of the prominence pillar is evident in the observations in Figure~\ref{f:ondisk_context} of about $\sim40$\arcsec in the plane of the sky (i.e., the lower limit). In the model we used $\Delta z=0.4$ for the height of the pillar, with the coordinates normalization of $0.1R_s$ the vertical extent of the model prominence pillar is $0.04R_s=28$ Mm in agreement with observations. Clearly, the extent of the observed low temperature prominence pillar material of $\sim28$ Mm is much longer than the scale height for the $10^4$ K prominence material of $\sim 0.6$ Mm. Thus, one does not expect to see the cool material in gravitational equilibrium at these heights in a field-free or in purely vertically directed magnetic field region, and the prominence material must supported by a horizontal magnetic field component.

The boundary conditions at $x=x_{min}$ and $x=x_{max}$ are line tied, and the other boundary conditions are open except for the lower boundary of the prominence pillar ($z=z_{min}=1$). In order to launch the NFM waves at the lower pillar boundary, the following time-dependent boundary conditions are applied on the $V_z$ velocity component:
\begin{eqnarray}
&&V_z(t,x,y,z=z_{min})=\frac{V_d}{2}\left[1+{\rm cos}(\omega t)\right]e^{-[(x-x_0)/s_x]^4-[(y-y_0)/s_y]^2}.
\label{Vzt0:eq} 
\end{eqnarray}
 Motivated by the observed wave propagation primarily inside the pillar as evident in Figure~\ref{f:2012Feb14CaIIimg} and the related  animations, the source of the wave flux is set to be maximal in the center of the prominence pillar by using the parameters $x_0=y_0=0.0$, $s_x=0.10$, and $s_y=0.15$; the amplitude $V_d$ controls the nonlinearity. The density and magnetic field perturbations are computed by zero-order interpolation from the interior of the computational domain, whereas the transverse velocity components are set to zero at the boundary. This results in periodic perturbations of the magnetic field, density, and velocity $V_z$ that inject NFM waves into the prominence pillar structure. In Table~\ref{param:tab}, we provide the values of $V_d$, $\omega$, $B_{x,0}$, and $B_{y,0}$ for the nine modeled cases in the present study. The results of Case~0 without waves are provided for reference.

\begin{table}
\caption{Parameters of the numerical 3D MHD models of prominence pillars with waves and flows. The velocity amplitude is given in units of $V_A$, the frequencies in $\tau_A^{-1}$, and the magnetic field strength in Gauss.}
\centering
\hspace{1cm}\begin{tabular}{cllccc}
\hline
Case \# & $V_d\ [V_A]$ & $\omega$ $[ \tau_A^{-1}]$ &  $B_{x,0}$ [G] &   $B_{y,0}/B_0$ \\ \hline
     0 &   0        &  - & 10 &  0 \\
        1  &  0.01            &  5.28        &  10 &  0 \\
                2 &  0.01            &  12.56        &  10 &  0 \\
        3 &  0.02               &  12.56         &    10     &     0  \\
       4 &      0.01         &    5.28    &           10 &     0.1 \\
      5 &        0.01        &    12.56     &       10    &  0.1 \\
      6 &        0.02        &    12.56     &       10    &  0.1 \\
      7 &        0.02        &   12.56     &              10   & 0.2\\
      8 &        0.01         &  6.28     &             20    & 0.1 \\
\hline
\hline
\end{tabular}
\\ 
\label{param:tab}
\end{table}

\section{Numerical Results} \label{num:sec}
\begin{figure}[ht]
\centerline{
\includegraphics[width=\linewidth]{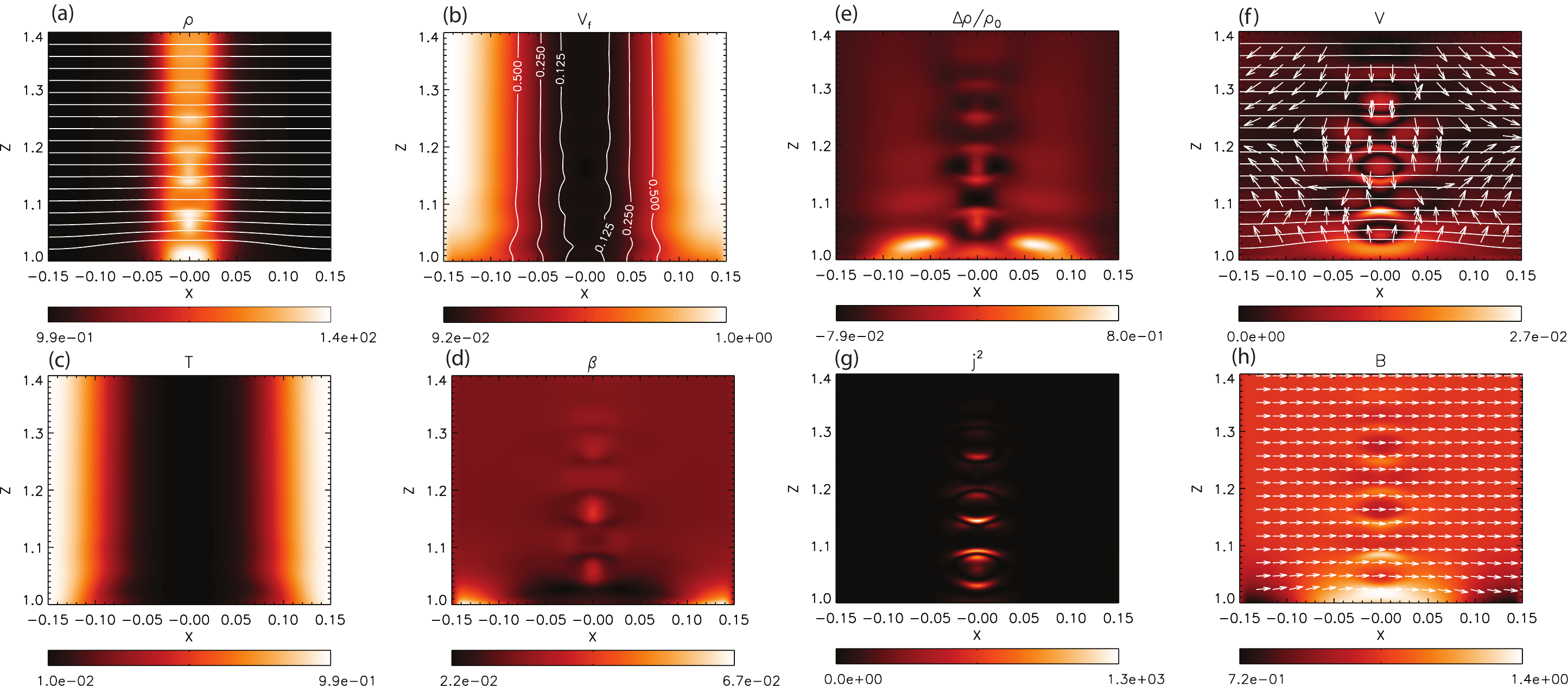}
}
\caption{The variables in the $x$-$z$ plane at $y=0$, $t=3.14$ with $V_d=0.02$, $\omega=12.56$ for Case 3. (a) $\rho$ with overlaid magnetic field lines. (b) The normalized fast magnetosonic speed $V_f$ magnitude with several isocontours, (c) $T$, (d) $\beta$, (e) $\Delta \rho/\rho_0$, (f) $V$ magnitude with arrows indicating the local direction of the velocity vectors, (g) $j^2$, (h) $B$ magnitude with arrows indicating the local magnetic field direction. An animation of panels (a) and (f) is available online. The video shows the density structure in the $x$-$z$ plane (left) with over-plotted field lines, and the corresponding velocity maps. The duration of the animation is 3.21 in normalized time units in the 5 s video.}
\label{nTVfbeta_jB_xz:fig}
\end{figure}

Here we present the results of the 3D MHD modeling of the NFM waves in the prominence pillar for the parameters given in Table~\ref{param:tab}. In order to explore the details of the waves, we first show in Figure~\ref{nTVfbeta_jB_xz:fig} the results in the $x$-$z$ plane cut at $y=0$ for Case~3 at $t=3.14\tau_A$. This prominence pillar is embedded in a uniform horizontal (potential) magnetic field modeled as described in Equation~\ref{B0:eq} with $B_{y0}$. The NFM waves are launched by the time-dependent velocity source ($V_z$, Eq.~\ref{Vzt0:eq}) at the lower boundary with amplitude $V_d=0.1V_A$ and frequency $\omega=12.56$ localized at the center of the prominence pillar. The waves propagate inside the pillar with nonlinear effects evident in non-modal structure of the oscillations, i.e., asymmetric and sharp peaks in the variables. The nonlinear wave pressure displaces the magnetic field lines upwards, as is evident in this figure, affecting the temperature, magnetosonic speed, and plasma $\beta$ structure. The details of the perturbations due to the waves are particularly clear in the density contrast, $\Delta\rho/\rho_0$. The prominence pillar acts as a leaky waveguide \citep{Cal86} for the NFM waves, as the magnetosonic speed $V_f$ is smaller by an order of magnitude inside the prominence pillar compared to the outside (coronal) region. The small leakage of the wave is most apparent in Figure~\ref{nTVfbeta_jB_xz:fig}e as the periodic density perturbations outside of the prominence pillar. The squared magnitude of the current density, $j^2$, shows the regions of enhanced currents that lead to Ohmic dissipation ($j^2/S$ in normalized units) associated with the NFM waves. The velocity components in the $x$-$z$ plane are shown Figure~\ref{nTVfbeta_jB_xz:fig}f, where the arrows indicate the local direction (not magnitude) of the velocity vectors and the magnitude $V$ is color-shaded as indicated by the color bar. The corresponding magnetic field in the $x$-$z$ plane is shown in Figure~\ref{nTVfbeta_jB_xz:fig}h. The dominant $B_x$ component is evident, along with the perturbations in the magnetic field magnitude $B$ due to the fast magnetosonic waves and the nonlinear wave pressure effects within the base of the prominence pillar.

\begin{figure}[ht]
\centerline{
\includegraphics[width=1.1\linewidth]{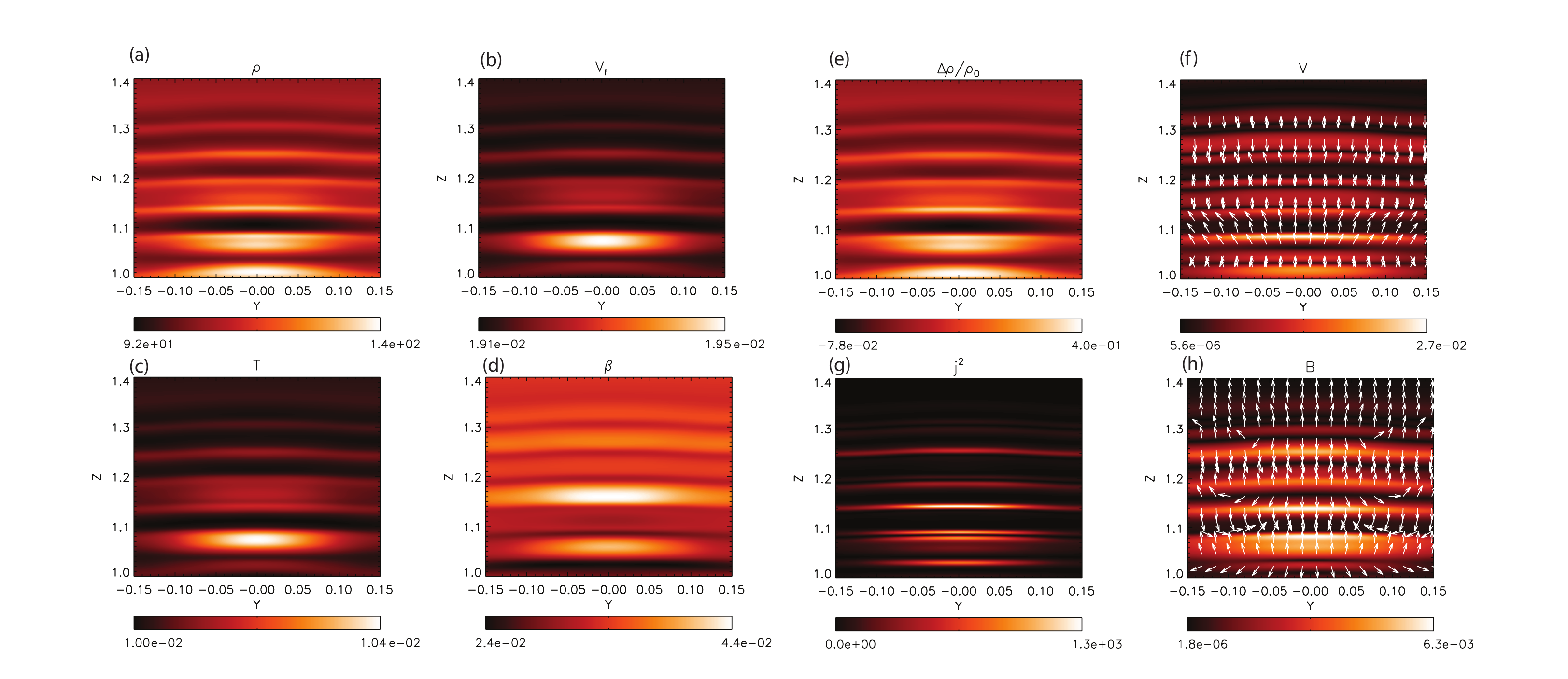}
}
\caption{The variables in the $y-z$ plane at $x=0$, $t=3.14$ for Case 3. (a) $\rho$, (b) $V_f$, (c) $T$, (d) $\beta$, (e) $\Delta \rho/\rho_0$, (f) $V$ with arrows showing the direction of the velocity, (g) $j^2$, (h) $B$ with arrows showing the direction of the magnetic field.}
\label{nTVfbeta_jB_yz:fig}
\end{figure}

In Figure~\ref{nTVfbeta_jB_yz:fig} we show the variables in a cut along the prominence pillar axis, i.e., in the $y$-$z$ plane at $x=0$ in the high-density, low-temperature (relative to coronal values) region at time $t=3.14\tau_A$. The effects of the NFM waves generated by the time-dependent boundary conditions in $V_z$ are evident. In particular, the density perturbations are in phase with the magnetic field perturbations, as seen by comparing the panels in Figure~\ref{nTVfbeta_jB_yz:fig}e and h, as expected for the fast magnetosonic waves. The magnitude of the waves is largest in the center of the pillar due to the form of the wave source in Equation~\ref{Vzt0:eq}, as well as to the waveguide trapping of the wave flux. The squared current magnitude $j^2$ is shown in Figure~\ref{nTVfbeta_jB_yz:fig}g, where the larger currents are associated with the wave fronts and are regions of higher Ohmic dissipation (therefore also affecting the temperature). The directions of the perturbations in $V$ and $B$ in the $y$-$z$ plane are shown in Figure~\ref{nTVfbeta_jB_yz:fig}f and h. The waves propagate in the $z$ direction since $V_f$ is nearly uniform in the $y$-$z$ plane, with very small perturbations due to the waves (note the intensity range in the color bar of Figure~\ref{nTVfbeta_jB_yz:fig}b).

The temporal evolution of the variables at a point in the center of the prominence pillar at x=0, y=0, z=1.2 for Cases~ 0-3 is shown in Figure~\ref{dbvnt_prom_vd0.01-0.02fd5.28-12.56B10_by00.0:fig}. The three components of the velocity and magnetic field perturbations (with respect to $\mbox{\bf B}_0$) and the change in density and temperature are displayed. The difference between Case~1 and Case~2 is the increase of wave frequency by a factor of 2.4, while in Case~3 the amplitude of the velocity at the boundary is increased by a factor of 2 with respect to Cases~1 and 2. Evaluating the propagation speed of the NFM waves from the animations for Cases 2 and 3, we find that they travel close and slightly above (5\%-15\%) the theoretical linear fast magnetosonic speed. The speedup is larger for the higher amplitude waves suggesting nonlinearity effect \citep[see,][and references therein]{OD97}. The nonlinearity of the fast magnetosonic waves is evident primarily in their non-sinusoidal temporal structure, which shows evidence of steepening. This nonlinear effect is more evident in the low-frequency waves and in the large-amplitude waves, where the wave peaks are sharper than the troughs due to steepening. The magnetic field perturbations show secular growth of the amplitude, an indication of nonlinear wave pressure effects on the background magnetic structure. The density perturbations show an oscillatory increasing trend, whereas the temperature perturbations are small. 

The case without without waves (Case~0) shows the evolution of the background state in the center of the pillar due to the gravitational settling of the initial state. One can estimate the magnetic field change expected for the given density increase of $\sim$5\% due to the gravitational settling.  This corresponds to a magnetic pressure change that is 5\% of $\beta$, or about 0.4\%, which equals to the value of $\Delta B/(2B_0)$.  Therefore, the estimated $\Delta B/B_0 =$0.8\% = 0.008 is consistent with the magnitude of the changes shown in Figure~\ref{dbvnt_prom_vd0.01-0.02fd5.28-12.56B10_by00.0:fig} in the field component plotted there for Case~0 (green curve). It is evident that the small velocity $V_z$ readjustment exhibits an initial oscillatory evolution due to the effect of gravity, followed by a nearly constant downward velocity $V_z\approx-0.003V_A$ corresponding in physical units to about $-2$ km s$^{-1}$. We investigated the effects of diffusion by repeating Case~0 with higher ($S=10^4$) and lower ($S=10^6$) resistivity. In the latter case the spatial resolution was doubled in each direction ($514^3$) compared to other runs. We find that in the case with $S=10^4$ the small down flow velocity increases by $30\%$. However, in the reduced resistivity, high resolution run with $S=10^6$, the asymptotic down flow speed remains nearly the same as in the case with $S=10^5$, where the density structure shows slight compression and broadening of the lower part of the pillar. The diffusion of cold prominence material through the supporting magnetic field is expected in real prominences in qualitative agreement with the present model {\bf for higher resistivity case}, since the material is partially ionized \citep[for example, see][]{Gil02,Kho14}, and where the frozen in condition breaks down due to finite resistivity \citep{Low12,LE14,JK21}, see the review by \citet{Gib18}. While in the MHD model the down flow velocity {at lower resistivity is due to compressive effects, we find that this velocity is  small compared to the phase speed of the fast magnetosonic waves and therefore has no significant effect on the wave propagation.
\begin{figure}[ht]
\centerline{
\includegraphics[width=\linewidth]{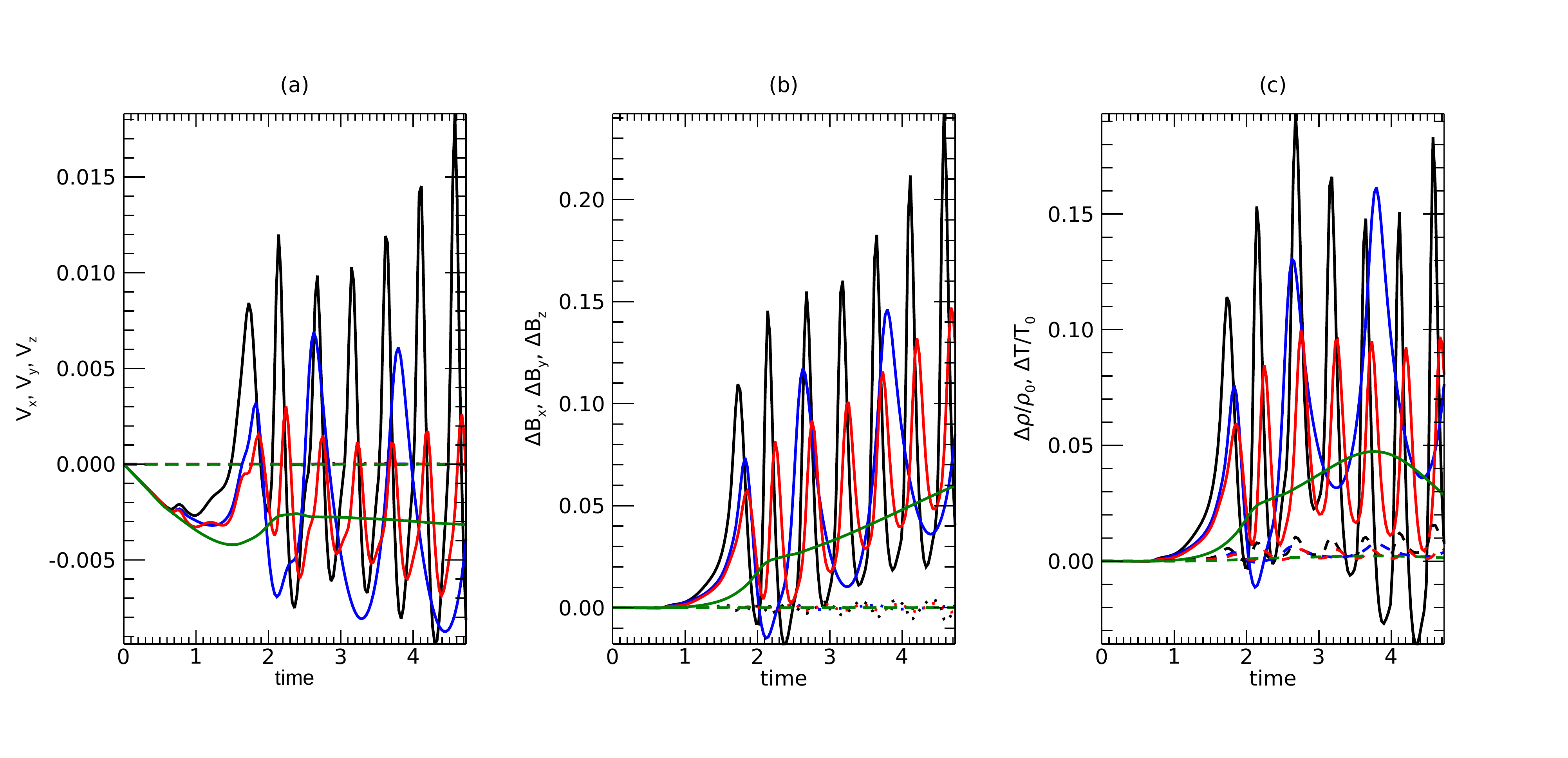}
}
\caption{Temporal evolution of the variables in the center of the prominence pillar for Case 1 (blue) with $V_d=0.01$, $\omega=5.26$, Case 2 (red) with $V_d=0.01$, $\omega=12.56$, and Case 3 (black) with $V_d=0.02$, $\omega=12.56$. (a) Velocity components $V_z$ (solid), $V_y$ (short dashes), $V_x$ (long dashes). (b) Magnetic field component perturbations $\Delta B_x$ (solid), $\Delta B_y$ (short dashes), $\Delta B_z$ (long dashes). (c) Changes in density $\Delta \rho/\rho_0$ (solid) and temperature $\Delta T/T_0$ (long dashes) normalized by the respective initial values $\rho_0$ and $T_0$. The case without waves (Case 0) is shown (green) for reference. Times are in units of $\tau_A$.}
\label{dbvnt_prom_vd0.01-0.02fd5.28-12.56B10_by00.0:fig}
\end{figure}

The structure of the magnetic field and density perturbations due to the NFM waves for Case~2 is shown in Figure~\ref{B3dn3d_by0:fig}. In the present model the initial state was the result of Case~0 without waves, where slight dips form in the magnetic configuration of the pillar due to the effects of reduced gravity. The figure and the animation show the magnetic field lines and the density isocontours in the domain at $t=5.8\tau_A$. A small lifting of the magnetic field lines by the wave pressure is evident  mostly in the lower region of the pillar, while the small gravitational dipping of the field lines is most evident in the upper part of the domain{ in the initial state, reduced at later times due to the effects of wave pressure. Isocontours of density indicate the locations of the propagating compressions due to the guided NFM waves, while further details of the wave propagation are exhibited in the animations provided. 
\begin{figure}[ht]
\includegraphics[width=0.43\linewidth]{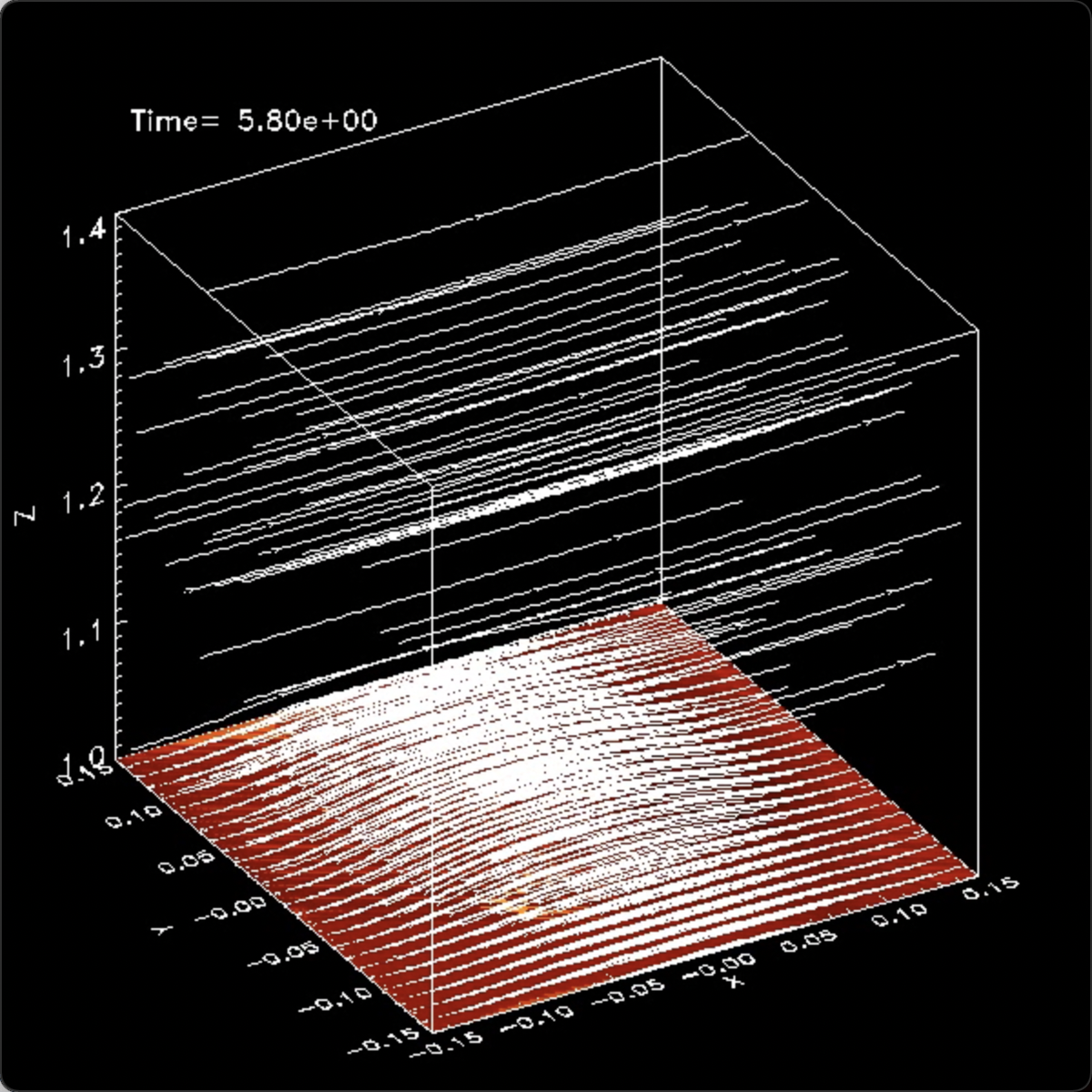}
\includegraphics[width=0.5\linewidth]{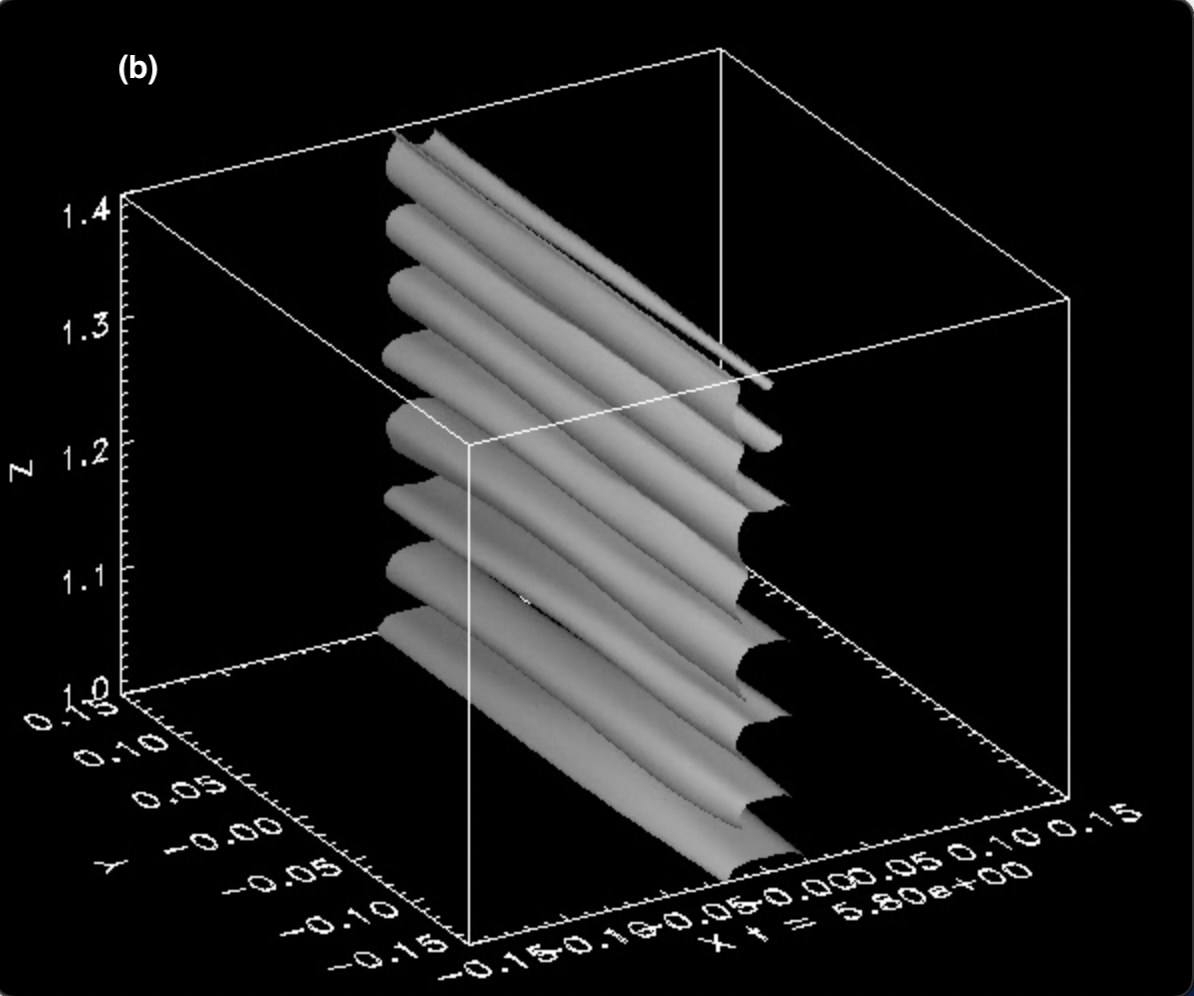}
\caption{The results of the 3D MHD model in Case~2. (a) Magnetic field lines and (b) density isocontours due to the propagating NFM waves in the domain at $t=5.8\tau_A$. An animation of this figure is available online. The duration of the animation is 3.6 in normalized time units that shown in 2 s video.}
\label{B3dn3d_by0:fig}
\end{figure}

The effects of non-potential, non-force-free magnetic fields on the propagation of the fast magnetosonic waves in the prominence pillar are explored in Cases~4-8. The form of the background magnetic field is given by Equation~\ref{B0:eq}. The parameters of Cases~4-6 are the same as in Cases~1-3 except that $B_{y,0}=0.1$. In Case~7 we consider $B_y=0.2$, with the other parameters as in Case~3, and in Case~8 we consider $B_0=20$ G, with the rest of the parameters the same as in Case~6. These results are discussed below.

In Figure~\ref{nTVfbeta_jB_by0.1_xz:fig} we show the variables in the $x$-$z$ plane at $y=0$ for Case~6 with $B_{y0}=0.1$ at $t=3.03\tau_A$. The NFM wave structure is most evident inside the prominence pillar in the relative density compressions, $\Delta\rho/\rho_0$, but also is seen in the variability in $\rho$, $\beta$, $j^2$, and the velocity and magnetic field magnitudes. Comparing $\Delta\rho/\rho_0$ to Case~3 (Fig.~\ref{nTVfbeta_jB_xz:fig}e), we find that the relative magnitude of the leakage in the $x$ direction is reduced. The effects of the $x$ component of the Lorentz force in compressing the prominence pillar density are seen by comparing the structure of $\rho$ to the initial state in Figure~\ref{B0_xy:fig} and to $\rho$ in Case~3 shown in Figure~\ref{nTVfbeta_jB_xz:fig}a. The apparent half-width is reduced by about $30\%$ in the present case. 
\begin{figure}[ht]
\centerline{
\includegraphics[width=\linewidth]{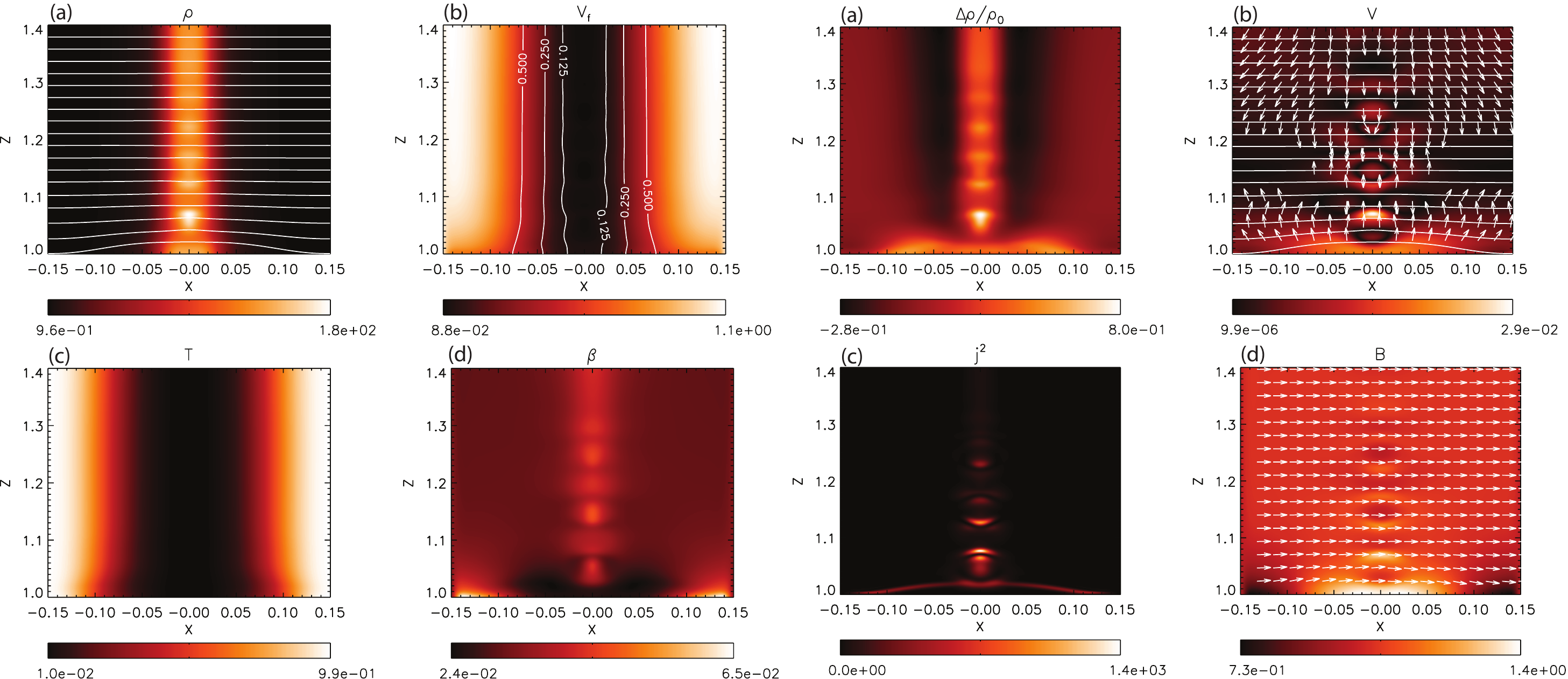}
}
\caption{The variables in the $x$-$z$ plane cut at $y=0$, $t=3.03\tau_A$ for $V_d=0.02$, $\omega=12.56$ (Case 6). (a) $\rho$ with field lines indicated with white lines. (b) $V_f$ magnitude with several isocontours. (c) $T$. (d) $\beta$. (e) $\Delta \rho/\rho_0$. (f) $V$ magnitude with arrows showing the local direction of the velocity vectors. (g) $j^2$. (h) $B$ magnitude with arrows showing the local direction of the magnetic field vectors (dominated by $B_x$).}
\label{nTVfbeta_jB_by0.1_xz:fig}
\end{figure}

Figure~\ref{nTVfbeta_jB_by0.1_yz:fig} shows the variables in the $y$-$z$ plane at $x=0$ for the case with $B_{y0}=0.1$ (Case~6) at $t=3.03\tau_A$. The refraction of the wave fronts of the NFM waves due to the effect of the $B_{y0}$ magnetic field component is apparent by comparison with Figure~\ref{nTVfbeta_jB_yz:fig}. The wave structure is evident in the density and magnetic field perturbations, as well as in the corresponding current perturbations. In this magnetic configuration, the waves leak significantly out of the prominence pillar through the side boundary at $y=y_{max}$, decreasing the wave energy flux in the center of the pillar, whereas in the uniform magnetic field case, the main leakage takes place through the top of the prominence pillar ($z=z_{max}$) with open boundary condition.
\begin{figure}[ht]
\centerline{
\includegraphics[width=\linewidth]{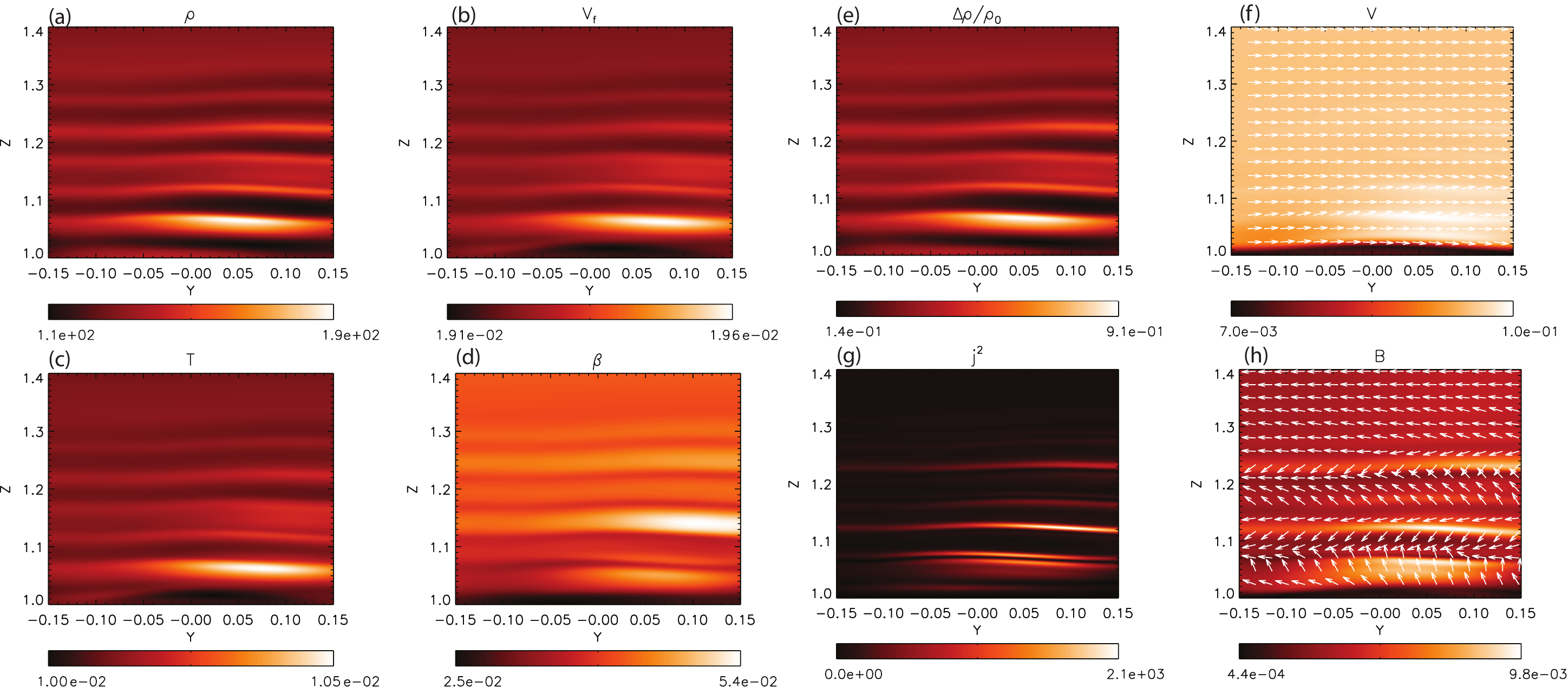}
}
\caption{Variables in the $y$-$z$ plane at $x=0$, $t=3.03$ for Case 6. (a) $\rho$. (b) $V_f$. (c) $T$. (d) $\beta$. (e) $\Delta \rho/\rho_0$. (f) $V$. (g) $j^2$. (h) $B$.}
\label{nTVfbeta_jB_by0.1_yz:fig}
\end{figure}

The cut in the $x$-$y$ plane (i.e., the solar `disk' view) of the model shows the structure of the prominence pillar density, temperature, fast magnetosonic speed, $\beta$, $j^2$, {\bf B}, and {\bf V} at height $z=1.2$ at $t=3.03\tau_A$ for Case~6. The  effects of the upward propagating NFM waves are evident. In the $x$-$y$ plane, the waves are most clear in $\Delta\rho/\rho_0$, $j^2$, $\beta$, and the magnitudes $B$ and $V$. The $y$ dependence of the wave structure is affected by both the driving source and the wave refraction due to the $B_{y0}$ component of the background magnetic field. It is evident from $\Delta\rho/\rho_0$ that the leakage of the wave is significant in the density compressions outside the prominence pillar region (i.e., $|x|>0.05$). The Lorentz force generates a compression of the density primarily in the $x$ direction, with small magnetic field and density compression in the $y$ direction, as can be seen from the density and velocity structures in the $x$-$y$ plane.
\begin{figure}[ht]
\centerline{
\includegraphics[width=\linewidth]{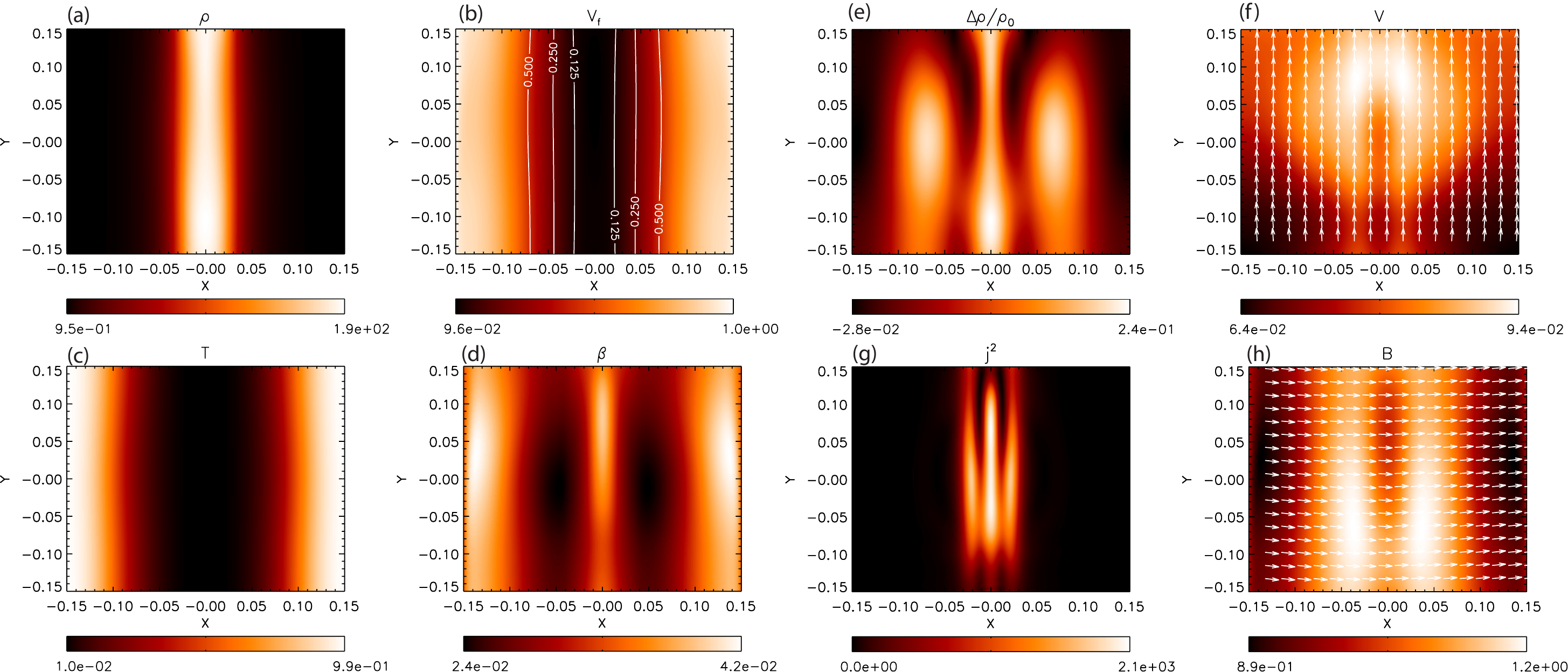}
}
\caption{Variables in the $x$-$y$ plane at $z=1.2$, $t=3.03$ for Case 6. (a) $\rho$. (b) $V_f$ with several isocontours. (c) $T$. (d) $\beta$. (e) $\Delta \rho/\rho_0$. (f) $V$ magnitude with direction arrows. (g) $j^2$. (h) $B$ magnitude with direction arrows.}
\label{vfbeta_xy_prom_vd0.02fd12.56B10_by00.1:fig}
\end{figure}

The temporal evolution of the variables for Cases~4-6 in the center of the prominence pillar at x=0, y=0, z=1.2 are shown in Figure~\ref{dbvnt_prom_vd0.01fd5.28-12.56B10_by00.1:fig}. Evidently, the non-force-free magnetic field introduces flows due to the Lorentz force that lead to compression in the prominence pillar, and corresponding increases of magnetic field strength and density that disrupt the initial gravitational equilibrium. In particular, it is evident that the $V_y$ component has similar accelerated evolution in Cases~4-6, with weak dependence on the properties of the fast magnetosonic waves. Thus, the effects of the Lorentz force in the non-force-free field in introducing mass flows self-consistently becomes evident. The flow accelerates during the simulated time, exceeding 10\% of the Alfv\'{e}n speed (about 70 km s$^{-1}$ with the present normalization) by the final modeled time.
\begin{figure}[ht]
\centerline{
\includegraphics[width=\linewidth]{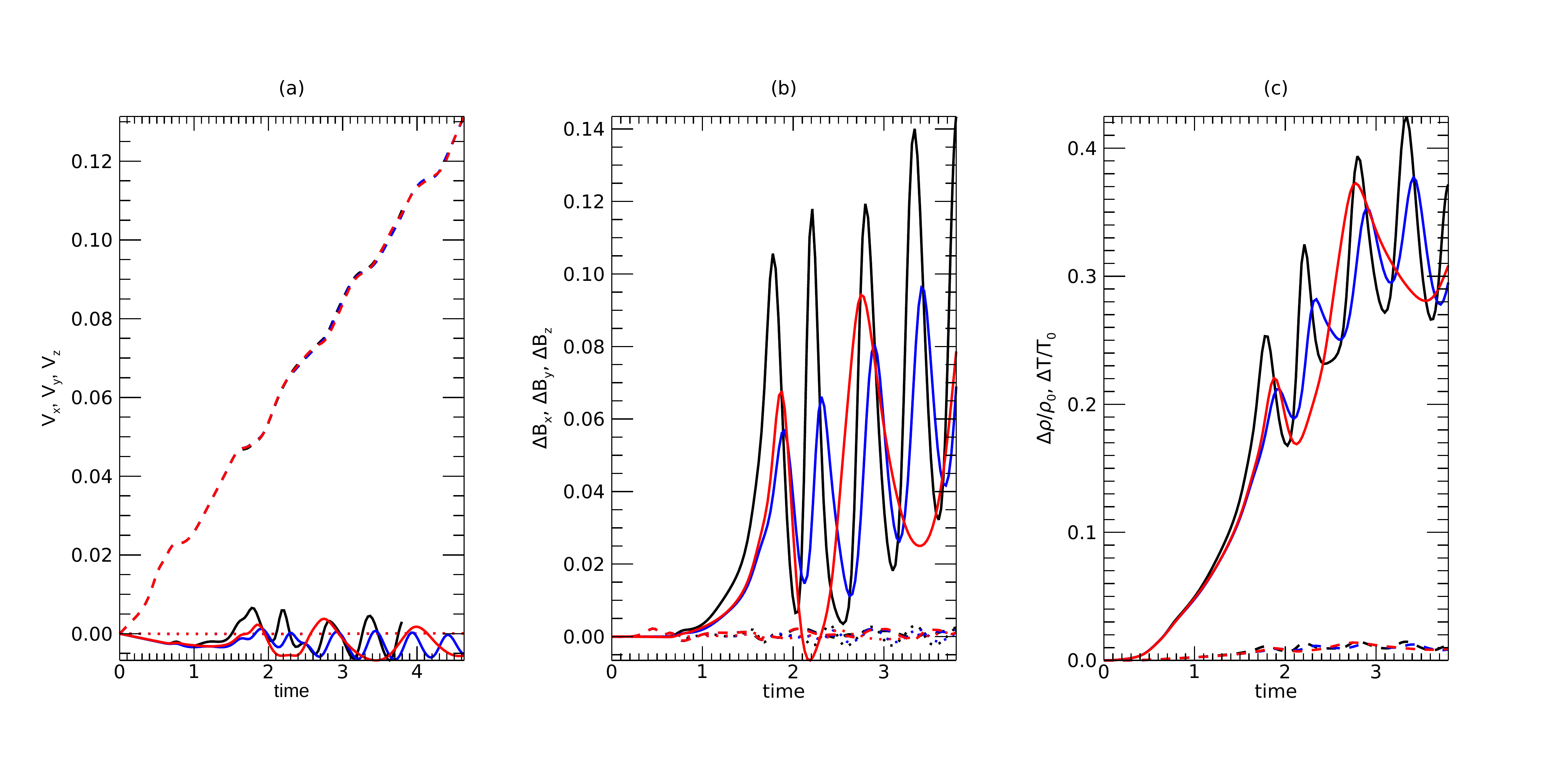}
}
\caption{The temporal evolution of the variables in the center of the prominence pillar for Case 4 (red) with $V_d=0.01$, $\omega=5.28$,  Case 5 (blue) with $V_d=0.01$, $\omega=12.58$, and Case~6 (black) with $V_d=0.02$, $\omega=12.58$. (a) Velocity components $V_z$ (solid), $V_y$ (short dashes), $V_x$ (long dashes). (b) Magnetic field component perturbations $\Delta B_x$ (solid), $\Delta B_y$ (short dashes), $\Delta B_z$ (long dashes). (c)  Changes in density $\Delta \rho/\rho_0$ (solid) and temperature $\Delta T/T_0$ (long dashes) normalized by the respective initial values $\rho_0$ and $T_0$. Times are in units of $\tau_A$.}
\label{dbvnt_prom_vd0.01fd5.28-12.56B10_by00.1:fig}
\end{figure}

The effect of increased Lorentz force on the structure of the prominence pillar and on the NFM waves is demonstrated in Case~7 with $B_{y0}=0.2$. As expected, the increased Lorentz force leads to more rapid and powerful compression of the prominence pillar than in Cases 4-6. This affects the properties of the background density structure of the pillar and also the NFM waves. In particular, the wave frequency has decreased due to the increased compression, mainly due to the increase in density and corresponding decrease in $V_f$ inside the prominence pillar. This also leads to the decrease of the velocity amplitude associated with the NFM waves inside the pillar for the fixed wave source at the boundary. 

In Case~8 we investigate the effects of increased magnetic field strength on the NFM waves by doubling the assumed magnitude of the magnetic field, $B_0=20$ G. This change with respect to previous cases results in a doubling of $V_A$ and a decrease of plasma $\beta$ by a factor of four. Since the velocity amplitude $V_d$ is in units $V_A$, the magnitude of the nonlinearity of the fast magnetosonic wave in Case~8 is essentially the same as in Case~4. Also, we note that $\tau_A$ is half the value in Case~8 compared to other cases, and that the value of $\omega$ in Case~8 is equal to the value in Case~7 when converted to rad s$^{-1}$. The main difference with respect to previous cases is the effect of the wave pressure, which is now four times larger in Case~8 and leads to correspondingly stronger density compressions. 

\section{Discussion and Conclusions} \label{dc:sec}

Recent high spatial and temporal resolution observations of a prominence pillar from Hinode/SOT in H$\alpha$ and Ca~II and IRIS in Mg~II show evidence of small-scale oscillations and propagating features associated with flows \citep{KOT18}. Analysis of Doppler shifts from Hinode/SOT \Ha\ and IRIS show red-wing/blue-wing contrasts that are consistent with propagating waves and flows on extended magnetic field lines \citep[e.g.,][]{OK20}. Here, we analyze additional observations of propagating small-scale oscillations in the Hinode/SOT Ca~II line and the blue wing of \Ha\ in a prominence on 2012 February 14. Using space-time plots and wavelet analysis, we find oscillations with typical periods of order minutes and wavelengths of order 1000-2000 km with sharp peaks indicative of nonlinear steepening, and we identify the propagating features as signatures of nonlinear fast magnetosonic (NFM) waves.

Motivated by past and recent observations of the small-scale oscillations, we developed an idealized 3D MHD model of a prominence pillar that focuses on the generation and propagation of NFM waves guided by observed properties. The advantage of the simplified 3D model is tractability of the wave features when the line-of-sight projection effects inherent to single-point plane-of-the-sky observations are removed. The 3D model reproduces the main physical properties of the prominence cool material embedded in background magnetic field and of the observed propagating small-scale features. 

The present model extends previous 2.5D MHD studies of the propagating NFM waves into more complex and realistic prominence structures by allowing 3D wave propagation and couplings. There is  evidence of nonlinear coupling of the NFM waves to other wave modes in the animation of the density structure, which shows secondary density compressions due to slow magnetosonic mode that appear to follow the compressions associated with NFM waves. There is also evidence of small amplitude Alfv\'{e}nic oscillation in the temporal signatures of the variables. However, we find that the main effects of nonlinearity of the waves are the steepening and the coupling between the NFM waves and the background pillar structure in the low-$\beta$ plasma.

We modeled eight cases and varied the main parameters of the waves in two types of magnetic field configurations (uniform potential and for the first time non-force-free) to provide insights on the effects of the various parameters on the generation and propagation of NFM waves. Evidence of velocity and magnetic shear is often observed in pre-eruptive prominences configurations \citep{Gib18}. Therefore, we modeled non-force-free field with magnetic shear that introduces large-scale flows, corresponding (aperiodic) compressions of the prominence pillar, and dynamic changes in wave propagation properties, all self-consistently. We find that the effects of the non-force-free  sheared magnetic field on the pillar structure and on the wave propagation are significant, even for relatively small magnitude of the shear-produced Lorentz force, due to the low-$\beta$ state of the prominence material. Thus, the effects of magnetic field shear on the NFM waves may affect the application of coronal seismology in prominence pillars.

The modeling results show qualitative agreement with the observed propagating oscillations with nonlinear steepening in the prominence pillar, as demonstrated in previous studies \citep{Ofm15a,OK20} and the present observational analysis. The 3D MHD results confirm further the interpretations of the observed propagating small-scale features in terms of NFM waves that are wave-guided in the cool material (low fast magnetosonic speed) of the prominence pillar region. From the model we find that the wave nonlinearity leads to secular changes in background magnetic field structure, density, and temperature due to the wave pressure, in addition to the wave steeping effects that affect the small-scale compressive structures. The low-frequency wave source leads to higher amplitude guided NFM waves than the high-frequency waves, due to lower leakage and dissipation compared to the high-frequency waves. 

Our study demonstrates the potential applications to the observed small-scale waves together with modeling for magnetic seismology of the prominence structure. One can apply coronal seismology by using the properties of the observed waves, such as wavelengths and periods to determine the phase speed of the waves. The relation between the phase speed and the magnetic field can be obtained from linear theory for linear waves in simplified geometry \citep[e.g.,][]{NV05}. For nonlinear  waves in more complex geometry the phase speed can be obtained from a 3D MHD model. . Finally, by comparing the theoretical/modeled phase speed with the observed phase speed and with the density and temperature information, one can determine the magnetic field in the pillar (taking into account possible plane of the sky (POS) projection effects). The details of magnetic geometry structure could be deduced from the observed direction of wave propagation where a 3D MHD model helps alleviate the POS observational ambiguity. The present model considers the nonlinearity in various idealized magnetic field geometry scenarios, and in the future more realistic 3D MHD wave models will include more detailed magnetic and density structure based on specific observations, thus improving the accuracy of coronal seismology method.


LO acknowledges support by NASA Cooperative Agreement 80NSSC21M0180 to The Catholic University of America. Resources supporting this work were provided by the NASA High-End Computing (HEC) Program through the NASA Advanced Supercomputing (NAS) Division at Ames Research Center. TAK and CRD were supported by NASA's H-ISFM program at Goddard Space Flight Center. Hinode is a Japanese mission developed and launched by ISAS/JAXA, with NAOJ as domestic partner and NASA and STFC (UK) as international partners. It is operated by these agencies in co-operation with ESA and NSC (Norway).  We are grateful to the late Ted Tarbell for helpful discussions concerning the Hinode/SOT data during our past collaborations. The Global Oscillation Network Group (GONG) Program, managed by the NSO, is operated by AURA Inc.\ under a cooperative agreement with the NSF. 

%

\vspace{5mm}
\facilities{Hinode/SOT, GONG}
\software{SolarSoft}

\end{document}